\documentclass[aps,prx,twocolumn,superscriptaddress, showspace,showkeys,floatfix]{revtex4}
\usepackage{amsmath}
\usepackage{bm}
\usepackage{amssymb}
\usepackage{graphicx}
\usepackage{epstopdf}
\usepackage{bmpsize}
\usepackage{epsfig}
\usepackage{color}
\usepackage{hyperref}
\usepackage{float}
\usepackage{braket}
\usepackage{lipsum}
\usepackage{soul}

\usepackage{graphicx}
\usepackage{dcolumn}
\usepackage{bm}
\usepackage{hyperref}
\usepackage[mathlines]{lineno}
\usepackage{amsmath}
\usepackage{amssymb}
\usepackage{graphicx}
\usepackage{epstopdf}
\usepackage{bmpsize}
\usepackage{epsfig}
\usepackage{color}
\usepackage{float}
\usepackage{braket}
\usepackage{lipsum}
\usepackage[toc]{appendix}
\usepackage{amsmath,mathtools,amssymb,tikz}

\begin{document}
\date{\today}

\author{H. Alaeian}
\thanks{These authors contributed equally to this work.}
\affiliation{Department of Physics \& Astronomy, Purdue University, West Lafayette, IN 47907, USA}
\affiliation{Department of Electrical \& Computer Engineering, Purdue University, West Lafayette, IN 47907, USA}

\author{M. Soriente}
\thanks{These authors contributed equally to this work.}
\affiliation{Institute for Theoretical Physics, ETH Zurich, 8093 Z\"urich, Switzerland}
\affiliation{Department of Physics, Harvard University, 17 Oxford Street, Cambridge, Massachusetts 02138, USA}

\author{K. Najafi}
\affiliation{Department of Physics, Harvard University, 17 Oxford Street, Cambridge, Massachusetts 02138, USA}
\affiliation{ IBM Quantum, IBM T.J. Watson Research Center, Yorktown Heights, NY 10598 USA}

\author{S. F. Yelin}
\affiliation{Department of Physics, Harvard University, 17 Oxford Street, Cambridge, Massachusetts 02138, USA}

\begin{abstract}
Driven-dissipative quantum many-body systems have been the subject of many studies in recent years. They possess unique, novel classes of dissipation-stabilized quantum many-body phases including the limit cycle. For a long time it has been speculated if such a behavior, a recurring phenomenon in 
non-linear classical and quantum many-body systems, can be classified as a time crystal. However, the robustness of these periodic dynamics, against quantum fluctuations is an open question. In this work we seek the answer to this question in a canonical yet important system, \textit{i.e.}, a multi-mode cavity with self and cross-Kerr non-linearity, including the fluctuation effects via higher order correlations. 
Employing the Keldysh path integral, we investigate the Green's function and correlation of the cavity modes in different regions. Furthermore, we extend our analysis beyond the mean-field by explicitly including the effect of two-body correlations via the $2^\textrm{nd}$-cumulant expansion. Our results shed light on the emergence of
dissipative phase transitions in open quantum systems and clearly indicate the robustness of limit-cycle oscillations in the presence of the quantum fluctuations.
\end{abstract}


\title{Noise-Resilient Phase Transitions and Limit-Cycles in Coupled Kerr Oscillators}

\maketitle

\section{Introduction}~\label{sec:intro}
Understanding the quantum phase transitions (QPT) of many-body systems and their time evolution are among the main goals of modern physics. Historically, phase transitions have been introduced and investigated in the thermodynamics limit of closed systems where they thermalize in the steady-state. In recent years, however, phase transitions and critical phenomena in driven-dissipative many-body quantum systems have emerged as a major field of research. Extensive studies have been exploring the new classes of quantum phase transitions in the non-equilibrium steady-state (NESS) of open quantum systems, on account of experimental realization of dissipative quantum simulators~\cite{Tindall2019,Petit2020, Soriente2021}.

Experimental platforms, such as cavity arrays, superconducting circuits, and exciton-polaritons, have put forward versatile testbeds to examine the interplay between (in)coherent drive, dissipation, and interaction on NESS phases. This includes the multi-stability and crystallization in driven-dissipative nonlinear resonator arrays~\cite{Rodriguez2016,Cao2016,Biondi2017,Collodo2019} and spins~\cite{Landa2020}, and synchronized switching in the arrays of coupled Josephson junctions~\cite{Leib2014}, to name a few. Limit-cycle (LC), \textit{aka} synchronization, is an intriguing phase of open quantum systems, where the dynamics traverses a closed trajectory in the phase space~\cite{Pikovsky2001,Strogatz2001,Eneriz2019}. 

So far, mean-field (MF) LC-phases have been predicted in a wide variety of open quantum systems including optomechanical resonators~\cite{Ludwig2008,Rodrigues2010,Qian2012,Nation2013,Lorch2014}, Rydberg lattices~\cite{Lee2011}, Bose–Hubbard arrays with cross-Kerr interactions~\cite{Jin2013,Jin2014,Heinrich2010,Buca2019,Alaeian2021}, Heisenberg lattices~\cite{Owen2017}, spin arrays~\cite{Chan2015}, Dicke model~\cite{Bhaseen2012,Gambetta2019}, and non-linear photonic crystal cavities~\cite{TAKEMURA2020}. Due to its underlying assumptions however, MF treatment does not include the quantum fluctuation effects, and only a few recent studies have investigated the notion of LC in the context of super-operator spectrum~\cite{Tindall2020}. 


\begin{figure}
    \includegraphics{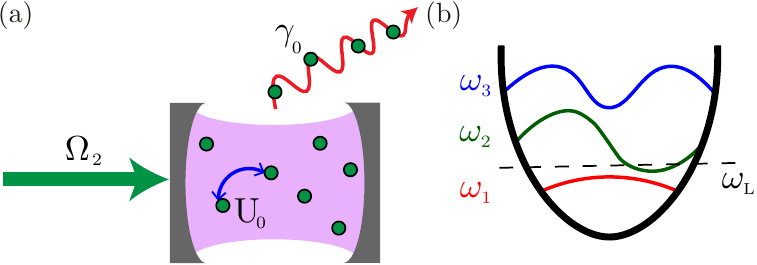}
    \centering
    \caption{(a) The schematics of a multi-mode spherical Fabry-Perot cavity containing a non-linear medium that leads to an effective photon-photon interaction at the rate $U_0$. (b) The bare cavity modes set by the cavity geometrical features, e.g. the mirrors radius of curvature and the cavity length.}
    \label{fig:schematics_intro}
\end{figure}

In this article, we ask the question of what are the elements that appear typically in the phase diagram of a dissipative non-equilibrium system. The focus is on directly experimentally realizable systems that allow in complexity at least for limit cycles. The theoretical methodology will be built upon typical tools, such as mean-field theory, and expanded as far as necessary to show and confirm the complex results. The story we tell in this article, thus, introduces first the physical system, which consists of a dissipative multi-mode cavity shown in Fig.~\ref{fig:schematics_intro} which is driven on one mode and connected to two side modes via a Kerr nonlinearity. Depending on the parameters, a complex phase diagram with various steady states, bistability, or limit cycles ensues. As a benchmark, this system is compared to the same system without the dissipation and a much simpler phase diagram that consists only of two types of steady-state. 

This is a complex interacting system where mean-field theory is not automatically expected to provide the correct results. We therefore support and correct the results using two different methods, (i) employing a thermodynamic-limit Keldysh formalism and (ii) including higher-order correlations. Both these methods confirm the phase diagram and provide the quantitative corrections.
The results and the methods presented in this work enable one to study the quantum properties of NESS beyond the semi-classical approximation and shed light on the emergence of dissipative phase transitions in open quantum systems. 

The paper is structured as follows. In section~\ref{sec:formalism} we introduce our model of a driven three-mode cavity with self and mutual Hubbard-type interaction subject to single-photon (coherent) drive. When the cavity-environment can be described by a Markovian process, the joint density matrix of the cavity evolves according to a Lindblad master equation. In section~\ref{sec:closed_system_results} we describe the physics of the closed system which we later use as a benchmark to highlight the profound effects of dissipation. Section~\ref{sec:results} presents the numerical results of an attractive interaction and details the MF phase diagram, explains the uniform and multi-stable regions for the pumped mode, and the emergence of the LC-phase for un-pumped side modes. Moreover, we elaborate on the spectral features of various phases obtained from Keldysh Green's functions, and showcase dissipation-stabilized cross-overs in the open system. The beyond-MF results employing the $2^\textrm{nd}$-cumulant approach is presented in sub-section~\ref{sub-sec: 3-mode cumulant expansion} where we compare the Gaussian approximation of the cavity states with their corresponding MF results. The implications of quantum correlation on LC is detailed in this sub-section. 
In section~\ref{sec:Keldysh} we briefly summarize our Keldysh formalism and the relevant Green's function presented in this work. Finally, section~\ref{sec:conclusion} concludes our results and presents an outlook for follow-up works.
\section{The model}~\label{sec:formalism}
We consider the dynamics of a three-mode lossy cavity with self- and cross-Kerr non-linearity, \textit{i.e.} a Hubbard interaction, subject to a single-photon coherent drive and a single-photon loss~\cite{Alaeian2021}. In the laser rotated frame, the conserved dynamics of such system is given via the following Hamiltonian ($\hbar = 1$)
\begin{align}~\label{eq:3-mode Hamiltonian}
    \hat{H} = & \sum_{m=1}^3 \Delta_m \hat{a}_m^\dagger \hat{a}_m + \frac{U_0}{2}\left(\hat{a}_m^{\dagger 2} \hat{a}_m^2 \right) \notag \\
    & \quad + 2U_0 \left(\hat{a}_1^\dagger \hat{a}_2^\dagger \hat{a}_1 \hat{a}_2 + \hat{a}_1^\dagger \hat{a}_3^\dagger \hat{a}_1 \hat{a}_3 + \hat{a}_3^\dagger \hat{a}_2^\dagger \hat{a}_3 \hat{a}_2 \right) \notag \\
    & \quad + \left(U_0 \hat{a}_2^{\dagger 2} \hat{a}_1 \hat{a}_3 + \Omega_2 ~ \hat{a}_2^\dagger + \textrm{H.c.} \right)\,, 
\end{align}
where the $2^\textrm{nd}$-mode is subjected to a single-photon coherent drive at the rate of $\Omega_2$ and the frequency of $\omega_L$. $\hat{a}_m, \hat{a}_m^\dagger$ is the annihilation and creation operators of the $m^\textrm{th}$-cavity mode, respectively, $\Delta_m = \omega_m - \omega_L$ is the detuning of the $m^\textrm{th}$ cavity mode from the coherent drive, and $U_0$ is the interaction rate. 

For $U_0 \ge 0$ the system energy increases with increasing the particle number hence, a \emph{repulsive} interaction. Similarly, when $U_0 \le 0$ the system energy decreases with increasing the particle number and the interaction is \emph{attractive}.

In the presence of an incoherent single-photon loss with the rate of 2$\gamma_m$ from the cavity, the Markov-Born approximation leads to the following Lindblad dissipator for the cavity density operator
\begin{equation}~\label{eq:single-photon dissipator}
    \mathcal{D}(\hat{\rho}) = \sum_{m=1}^3 \gamma_m \left(2\hat{a}_m \hat{\rho} \hat{a}_m^\dagger - \{\hat{a}_m^\dagger \hat{a}_m , \hat{\rho} \} \right).
\end{equation}
The time evolution of the multi-mode cavity density operator $\hat{\rho}(t)$ is determined via the following master equation 
\begin{equation}~\label{eq:Liouvillian}
    \frac{d}{dt}\hat{\rho}(t) = - i [\hat{H} , \hat{\rho}] + \mathcal{D}(\hat{\rho}).
\end{equation}
In this work, we are interested in the long-time solution of this density matrix, where all the transient dynamics are over. This is obtained by numerically integrating the coupled equations of motions (EoM) [cf. Appendix~\ref{app:MF_EOM}] implemented by the $4^\textrm{th}$-order Runge-Kutta method.

When the driving laser is swept from the red to the blue detuning, the nonlinear dynamics sets in as a parametric amplification process and photons are created in the side modes as correlated pairs. As shown in Appendix~\ref{app:MF_EOM}, for an attractive interaction, a multi-stability region exists for the red-detuned driving fields satisfying $\Delta_2 \ge \sqrt{3} \gamma_2$. Similarly, for repulsive interactions a multi-stability phase exists for the blue-detuned coherent drives when $\Delta_2 \le - \sqrt{3} \gamma_2$. For some pumping rates and detuning, the amplification moves into a regime of self-sustained oscillations, \emph{aka} limit cycle, due to the non-linear self- and cross-Kerr coupling. In the next section, we detail the phase diagram of the aforementioned system assuming an attractive interaction.

\section{Closed system}\label{sec:closed_system_results}
To draw a preliminary understanding of our model we first describe the closed system. We begin by noting that the Hamiltonian of Eq.~\eqref{eq:3-mode Hamiltonian} is unchanged under the exchange of the un-pumped modes, $a_1 \leftrightarrow a_3$. Additionally, there are neither terms that mimic a single- nor a two-photon pump for these modes taken separately. Therefore, a symmetry-breaking due to population of modes $a_1$ and/or $a_3$ is only possible if both of them are populated at the same time. On the contrary, due to the single-photon pump of the $2^\textrm{nd}$-mode, the system features a phase transition as a function of the pump strength $\Omega_2$. We study the mean-field energy potential landscape, $\bar{H}_{3}$, for a zero occupation of the $1^\textrm{st}$ and $3^\textrm{rd}$-mode~\footnote{The study of the full $6-$dimensional energy functional goes beyond the scope of this work.}.
\begin{equation}
\label{eq:3mode_MF_landscape}
    \bar{H}_{3} = \Delta_2 |\alpha_2|^2 + \frac{U_0}{2}|\alpha_2|^4 + \Omega_2 \alpha_2^* + \Omega_2 \alpha_2\,,
\end{equation}
where $\alpha_2$ is the order parameter.
Without loss of generality, we assume $\Omega_2$ to be real and find the potential extrema to be given by
\begin{align}
  & U_0 {\textrm{Re}{(\alpha_2)}}^3 + \Delta_2 \textrm{Re}{(\alpha_2)} + \Omega_2 = 0 \,,\\
  & \textrm{Im}{(\alpha_2)} = 0\,.
\end{align}
This leaves us with a cubic equation in $\textrm{Re}{(\alpha_2)}$.
We study the discriminant and find the boundaries between the regions with only one and three possible solutions
\begin{equation}
  \Omega_2 = \frac{2}{3\sqrt{3}}\sqrt{\frac{\Delta_2^3}{|U_0|}}\,,
\end{equation}
where we explicitly assumed $U_0<0$ in our case. 

In the parameter regime where three solutions are allowed, we find one \emph{low} (LP) and one \emph{high} population (HP) phase accompanied by an un-physical one, \textit{i.e.} imaginary eigen-frequencies of excitations. 

To understand the nature of the extrema of the potential, we find the excitation spectrum associated with the uniform case Hamiltonian~\eqref{eq:3mode_MF_landscape}~\cite{Soriente2021}. We obtain the eigen-frequencies
\begin{align}
    \omega_3^\pm = \pm\sqrt{\left(n_2 U_0 + \Delta_2\right)\left(3n_2 U_0 + \Delta_2\right)}\,,
\end{align}
with eigen-vectors
\begin{equation}
    \mathbf{v}_\pm = \begin{pmatrix} \frac{U\alpha_2^2}{\omega_3^\pm -\Delta_2 - 2Un_2} \\ 1 \end{pmatrix}\,,
\end{equation}
and their associated symplectic norms
\begin{equation}
    ds_\pm^2 = \left|\frac{U\alpha^2}{\omega_3^\pm -\Delta_2 - 2U n_2}\right|^2 - 1\,,
\end{equation}
where $n_2 = |\alpha|^2$. 

The symplectic norm describes the nature of a state of the system. Whenever all positive(negative) frequency eigenmodes have a positive (negative) symplectic norm the state of the is the ground state. On the other hand if a physical state of the system, i.e., a state with real excitation eigenmodes, has at least one excitation mode with negative (positive) norm for positive (negative) excitation frequency, then the state is an excited state of the closed system. Additionally, a positive symplectic norm is associated with particle-like processes where an external excitation is absorbed by the system whereas a negative norm underpins a hole-like process where an excitation in the system is destroyed~\cite{Soriente2021}. Finally, the symplectic norm helps identifying the so called negative-mass instabilities~\cite{Scarlatella_2019}.

\begin{figure}[h]
    \includegraphics[width=0.8\linewidth]{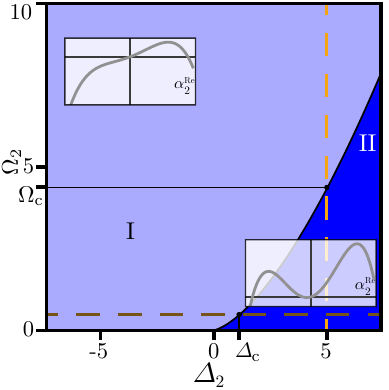}
    \centering
    \caption{Closed phase diagram of the three-mode uniform case. Two distinct regions are indicated by their respective mean-field energy potential ($\bar{H}_3$) landscape as a function of the real part of the cavity field, $\textrm{Re}{(\alpha_2)}$. The dark-blue region indicates the parameter regime where the LP phase is the ground state of the system and the HP represents a physically-allowed excited state. The LP phase disappears in the light-blue region where the HP is the only physical state.}
    \label{fig:3mode_closed_pd}
\end{figure}

In Fig.~\ref{fig:3mode_closed_pd}, we plot the three-mode mean-field energy potential of Eq.~\eqref{eq:3mode_MF_landscape}. Due to the single-photon pump of the $2^\textrm{nd}$-mode, the system features a phase transition as a function of the pump strength $\Omega_2$ between an LP and an HP phase. The phase diagram comprises of two qualitatively-distinct regions in parameter space with: I, the energy functional has only one clear extremum, the HP phase; II, three extrema exists including a ``proper'' LP phase ground-state, a saddle point, and a HP phase that reveals itself as a maximum, \textit{i.e.} an excited-state. 


\begin{figure}[h]
    \includegraphics[width=\linewidth]{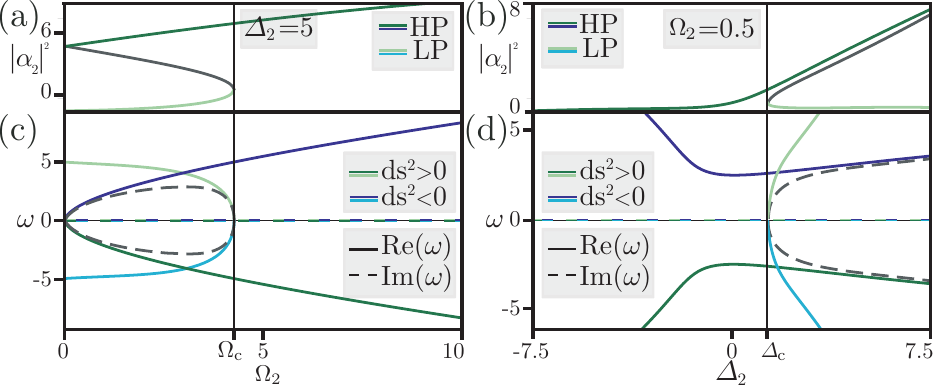}
    \centering
    \caption{(a),(b) The order parameter, $|\alpha_2|^2$, and (c)-(d) the excitation spectrum, $\omega$, on top of the LP (light hues) and HP (dark hues), along the orange and brown cut lines of Fig.~\ref{fig:3mode_closed_pd}, as a function of the pump strength, $\Omega_2$, and detuning $\Delta_2$, respectively. Real (solid) and imaginary (dashed) values with green [blue] hues encode the particle- [hole-]excitations, i.e. $ds^2>0$ [$ds^2<0$]. In (a), (c), at the $\Omega_{c}$ boundary between the light blue and dark-blue regions in Fig.~\ref{fig:3mode_closed_pd}, the LP ceases to exist and the only possible state is the excited-state HP. In (b), (d), at the $\Delta_{c}$ boundary, the LP appears as a possible state of the system. The HP always feature a norm swap with negative norm for the positive eigenmode. The gray lines indicate the third unphysical solution.}
    \label{fig:3mode_closed_eigen}
\end{figure}

In Fig.~\ref{fig:3mode_closed_eigen}, we plot the order parameter $|\alpha_2|^2$ and the excitation spectra on top of the possible solutions along the vertical orange cut (a,c) and the horizontal brown cut (b,d) of Fig.~\ref{fig:3mode_closed_pd}. In both cases the order parameter $|\alpha_2|^2$ has a finite value within the entire range but in region II, two distinct phases with different populations are possible. Besides, there is a third phase whose spectrum is fully imaginary and therefore not a physical state (gray lines in Fig.~\ref{fig:3mode_closed_eigen}(a),(b)). The HP exists throughout the entire phase diagram but always presents a negative (positive) symplectic norm with a positive (negative) eigenfrequency. Therefore, we confirm that it is indeed an excited-state of the closed system. Due to the presence of the negative interactions a ``true ground-state'' of the unbounded Hamiltonian is only possible in a limited region of the parameter space (see Fig.~\ref{fig:3mode_closed_pd}) and is identified with the LP phase, accompanied by a positive (negative) symplectic norm with a positive (negative) eigenfrequency~\cite{Soriente2020,Soriente2021}. 
\section{Open System}\label{sec:results}
%
For the open system, we determine the phase diagram by initializing the system in many randomized initial conditions and examining the dynamical stability of the end points using the Bogoliubov matrix spectrum [cf. Appendix~\ref{app:MF_EOM}]. In this respect, the phase boundaries can be thought of as dynamical phase transitions separating distinct long-time asymptotic behaviors~\cite{Heyl2013}. 

In addition to multi-stable phases, where the final state depends on the initial conditions, it is also possible to find regions of parameter space where no stable fixed point exists.
In such cases the system may be attracted to some time-dependent solutions such as limit cycles~\footnote{Note that in the rotated frame of these oscillation the limit-cycle phase can be considered as steady-state stagnation points.}, as found in other coupled nonlinear systems. In particular, we search for the complete set of stable attractors of the long-time dynamics, including the fixed points, the multistable coexistence phases, and time-dependent trajectories.
The possible steady-states for the evolution of the system under Eq.~\eqref{eq:Liouvillian} are phases where either the only populated mode is the pumped mode, $(\alpha_{n=2} \ne 0, \alpha_{n \ne 2} =0)$, called the \emph{uniform phase}, or phases where the side modes, $a_1,a_3$, are also populated and host a \emph{limit-cycle}. 

\begin{figure}[htbp]
\centering
\includegraphics[width=0.9\linewidth]{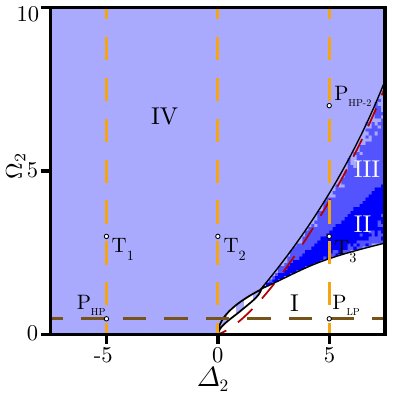}
\caption{\label{fig:phase diagram} The multi-mode open cavity phase diagram as a function of the pumped mode detuning $\Delta_2$ and the pumping rate $\Omega_2$. Phase transitions and crossover occur between a uniform low population phase (I, white) and a high population one (IV, light-blue). The dark-blue (II) and blue (III) regions indicate the parameter regime, where the low and high population phases are ``co-stable". Moreover, in the former a limit-cycle appears. The red-dashed line indicates the closed-system phase diagram boundary [cf. Fig~\ref{fig:3mode_closed_pd}]. The white-light-blue striped region, separating region I and IV, delimits an exceptional-point region and hosts a smooth crossover between the low and high population phases. Parameters are $\Delta_{1,3} = \Delta_2 \mp 1$ for the bare-cavity modes and $U_0 = -1$ for an attractive interaction, in terms of the cavity-mode decay.}
\end{figure}

In Fig.~\ref{fig:phase diagram} we plot the three-mode cavity open phase diagram. It comprises four different regions: (I) the white region that presents only one uniform stable steady-state which we identify as the LP phase, (II) the dark blue region of tri-stability between two uniform phases namely, the LP and the HP one, and a non-uniform solution where the un-pumped modes showcase a limit-cycle behavior, (III) the blue region of bi-stability between the LP and the HP uniform phases, and finally, (IV) the light blue region that presents a HP uniform phase. At the boundary between region (I) and (IV), denoted as the striped region, an exceptional-point appears and the LP and HP are smoothly connected. 

When compared to the phase diagram of the closed system [cf. Fig.~\ref{fig:3mode_closed_pd}], it becomes clear that the dissipation has a marked impact. It lifts the boundary of the closed phase diagram (red dashed line), stabilizes an excited state of the closed system, the HP, in region (IV), renders the LP as the only attractor of the dynamics in region (I), and leads to the emergence of limit-cycles.

\subsection{Steady-state}~\label{sub-sec:MF theo.}
In this section, we focus on different points of the open phase diagram in order to clearly expose the nature of the underlying attractors including their stability. As discussed in Appendix~\ref{app:MF_EOM}, the possible fixed points of the EoM are determined as $\frac{d}{dt}\braket{\hat{a}_m} = 0$ in Eq.~\ref{eq:MF 3-mode cavity}. 
Among the possible fixed points in the long-time limit, the semiclassical dynamics eventually evolves towards some stable fixed points for all initial conditions, known as the steady-states of the system.

We use the mean-field approximation, \textit{i.e.} we factorize higher order moments as $\braket{\hat{a}_m^\dagger \hat{a}_n} \approx \braket{\hat{a}_m^\dagger} \braket{\hat{a}_n}$, and obtain the equations of motion for the MF order parameters $\alpha_m$, with $m=1,2,3$. When $U_0 \Delta_2 \ge 0$ the EoM has one stable solution only as $(\alpha_{n=2} \ne 0, \alpha_{n \ne 2} =0)$, whereas when $U_0 \Delta_2 \le 0$, there might exist several stable steady-state solutions [cf. Appendix~\ref{app:MF_EOM}]. We self-consistently determine the steady-state of the system which, in general, may allow limit-cycle solutions. For these solutions, the long-time limit of the steady-state has the general form of $e^{it\omega_\textrm{LC}}$. Physically, the LC solutions and their associated frequencies ($\omega_\textrm{LC}$) can be thought as the frequency of the parametrically generated pair, i.e., the un-pumped modes, via the parametric Kerr process.

\begin{figure}[htbp]
\centering
\includegraphics{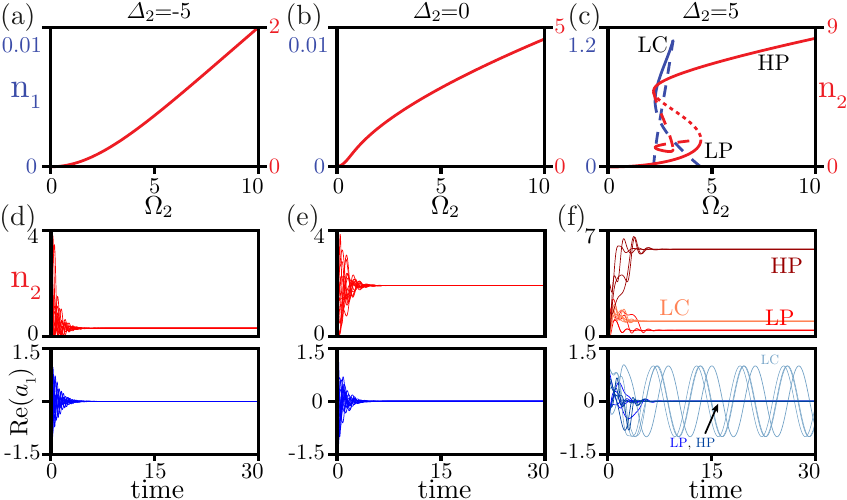}
\caption{\label{fig:3-mode MF} Cavity mode occupation vs. coherent pump rate ($\Omega_2$) for $\Delta_2 = -5, 0, +5$ in (a)-(c), respectively [cf. orange dashed cut-lines in Fig.~\ref{fig:phase diagram}]. In each panel, the red lines show the pumped mode population, $n_2$, and the blue lines show the un-pumped modes occupation, $n_1 = n_3$. Solid lines show the stable solutions while the dashed lines correspond to unstable branches. In (a) and (b), there is only a uniform stable steady-state (HP) whereas in (c) multiple stable steady-states are possible (LP, LC, HP). 
Panels (d)-(f) show the time evolution of the mean-field equations of motion for the corresponding detuning of (a)-(c) for 100 randomized initial conditions. In each panel the upper row shows the population of the pumped mode ($n_2 = |\alpha_2|^2$) and the lower row shows $\textrm{Real}(\alpha_1)$. In (d) and (e) the system always evolves towards a uniform steady-state with only the pumped mode populated (HP). In (f) there are three possible steady-states, two uniform phases LP, HP (red and brown) and a limit-cycle (LC, orange). In the limit-cycle case the side-modes show an oscillatory behavior at the frequency $\omega_\textrm{LC}$, as depicted with light blue traces in the bottom panel (f). All modes have the same decay rate $\gamma_0$ to which other rate parameters are normalized. The time is in units of $1/\gamma_0$. $U_0 = -1, \Omega_2 = 3$ in all temporal calculations. The side modes are equally-spaced around the pumped mode as $\omega_{1,3} = \omega_2 \mp 1$.}
\end{figure}

To draw a more comprehensive picture of the steady-state behavior of the system, in Figure~\ref{fig:3-mode MF}(a)-(c) we show the mean-field occupation of each cavity mode, $n_m=|\alpha_m|^2$, versus the pumping rate $\Omega_2$ at various detunings, $\Delta_2$ = -5, 0, +5 (vertical dashed lines in Fig.~\ref{fig:phase diagram}), for an attractive interaction of $U_0 = -1$, and bare cavity spacing as $\Delta_{1,3} = \Delta_2 \mp 1$.
The red lines show the behavior of the $2^\textrm{nd}$-mode while the blue lines correspond to the $1^\textrm{st}$- and the $3^\textrm{rd}$-mode populations. The solid and dashed lines signify the stable and unstable solutions, respectively. For $U_0 \Delta_2 \ge 0$, panels (a)-(b), there is only one stable steady-state, \textit{i.e.} the uniform phase HP, but when $U_0 \Delta_2 < \sqrt{3}$ we observe the regions of multi-stability and transitions to the non-uniform phases, Fig.~\ref{fig:3-mode MF}(c). 

In Fig.~\ref{fig:3-mode MF}(d)-(f) we show three exemplary MF time traces for randomized initial conditions corresponding to the points $T_{1,2,3}$, \textit{i.e.} $\Omega_2 = 3$, in Fig.~\ref{fig:phase diagram}. They are obtained from the direct integration of the EoM and illustrate the implications of a limit-cycle phase and its difference with the uniform one. For each detuning, the upper row shows the temporal behavior of $n_2 = |\alpha_2|^2$ while the lower panel shows the real part of the $1^\textrm{st}$-mode order parameter, \textit{i.e.} $\textrm{Re}(\alpha_1)$. 

For $T_1$ and $T_2$, corresponding to $\Delta_2 = -5,0$ respectively, there is only a uniform phase where $\alpha_{1,3} = 0$ and $\alpha_2 \ne 0$ [cf. Appendix~\ref{app:MF_EOM}]. Accordingly, in Fig.~\ref{fig:3-mode MF}(d),(e) we see that all time traces converge to only one non-zero value (HP) for the pumped mode and a zero value for the side modes, independent of the initial conditions.

On the other hand, for $T_3$ corresponding to $\Delta_2 = +5$, where a multi-stability is predicted [cf. Appendix~\ref{app:MF_EOM}], one can see that the time traces in the upper panel of Fig.~\ref{fig:3-mode MF}(f), converge to three different values for $n_2$ at LP (red traces), LC (orange traces), and HP (brown traces). From those three phases only the LC corresponds to a non-zero values for the $1^\textrm{st},3^\textrm{rd}$ modes (orange line in upper panel of the Fig.~\ref{fig:3-mode MF}(f)), where their order parameter $\textrm{Re}(\alpha_1)$ shows a periodic long-time behavior (ligh blue in the lower panel of Fig.~\ref{fig:3-mode MF}(f)).

\subsection{Uniform phase crossover}
\label{sub-sec:spectra}

\begin{figure}[h]
    \includegraphics[width=\columnwidth]{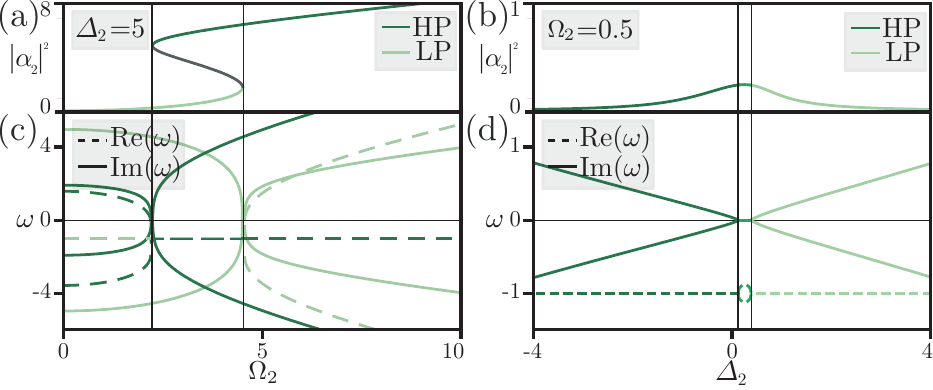}
    \centering
    \caption{(a,b) The order parameter, $|\alpha_2|^2$, and (c,d) real (dashed) and imaginary (solid) part of the excitation spectrum, $\omega$, on top of the LP (light green) and HP (dark green), along the vertical ($\Delta_2 = +5$) and horizontal ($\Omega_2 = 0.5$) cut lines in Fig.~\ref{fig:phase diagram}}
    \label{fig:3mode_open_eigen}
\end{figure}

To illustrate the dissipation effect in modifying the closed-system phases, in Fig.~\ref{fig:3mode_open_eigen}(a,c) and (b,d), we plot the population of the $2^\textrm{nd}$ mode, $|\alpha_2|^2$, (top) and the fluctuation eigenvalues (bottom) along the vertical ($\Delta_2 = +5$) and horizontal cuts ($\Omega_2 = 0.5$) of Fig.~\ref{fig:phase diagram}, respectively. In both cases, $|\alpha_2|^2$ behavior highlights a smooth cross-over from the HP to the LP between region IV and I. As can be inferred from the phase diagram, this transition traverses an exceptional point region where the imaginary part of the eigenvalues coalesce to zero and their real parts split, hosting over- and under-damped fluctuations.

\begin{figure}[h]
    \includegraphics{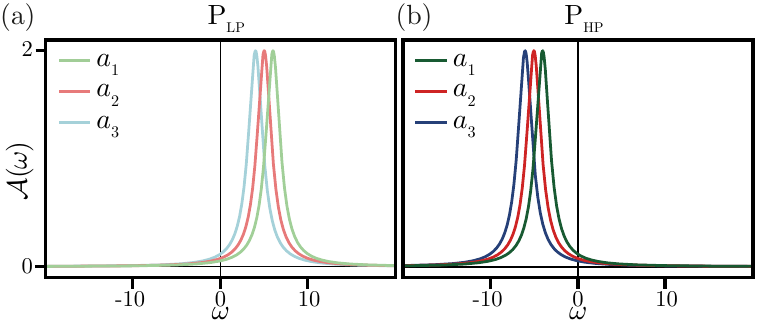}
    \centering
    \caption{Spectral function of the three-mode harmonic cavity uniform phase, for the points $P_\textrm{LP} = (\Delta_2 = +5, \Omega_2 = 0.5)$, (a), and $P_\textrm{HP} = (\Delta_2 = -5, \Omega_2 = 0.5)$, (b), [cf. Fig.~\ref{fig:phase diagram}]. The spectral function of $P_\textrm{HP}$ features a peak swap with respect to the one of $P_\textrm{LP}$. This signals the presence of a transition between particle- to hole-like physics going from region I to IV of the open phase diagram.}
    \label{fig:3mode_spectral_function_LP_HP}
\end{figure}

Using a Keldysh action approach we readily get access to the Green's function of the system and its associated dynamical observable. In Fig.~\ref{fig:3mode_spectral_function_LP_HP}, we compare the spectral functions, $\mathcal{A}(\omega) = -2\text{Im}[G^R(\omega)]$, of the three-modes in the LP (a) and HP (b) phases for different detuning and at fixed pump strengths $\Omega_2=0.5$, points $P_\textrm{LP,HP}$ in Fig.~\ref{fig:phase diagram}. From the left panel we see that the LP phase has the response of a ``ground-state", i.e., a positive (negative) peak at the positive (negative) frequencies [cf. Sec.~\ref{sec:closed_system_results}]. We note that peaks at negative frequencies are unresolved due to the scale resolution. On the other hand, the HP phase features a peak swap with a positive (negative) peak at the negative (positive) frequencies, a hallmark of a stabilized excited-state [cf. Sec.~\ref{sec:closed_system_results} and Fig.~\ref{fig:3mode_spectral_function} for more information]. We note that the two uniform phases have different occupations for the $2^\textrm{nd}$-mode, see Fig.~\ref{fig:3-mode MF} LP and HP. Nevertheless, their spectral functions highlight a more profound difference rather than just a different mode occupation. Interestingly, the peak swap is also present in the response of the empty side modes and we can think of it as being in presence of a normal phase to excited normal phase transition~\cite{Soriente2020}.

Even though the LP and HP phases are not described by the same order parameters, through the study of the dynamical responses we identified a peak swap between their spectral functions and we traced it back to the LP being the ground-state of the closed system and the HP an excited-state~\cite{Soriente2021,Ferri2021}. The dissipation stabilizes an excited-state, the HP, and lifts the boundaries of the closed system phase diagram changing its topography, see Sec.~\ref{sec:closed_system_results}. The spectral functions shown in Fig.~\ref{fig:3mode_spectral_function_LP_HP} belong to two points in the region I and IV of the open phase diagram, \textit{i.e.} where only one stable attractor exists. The spectral functions at these two points present a peak swap, signaling a particle- to hole-like physics transition, when going from the region I to region IV of the open phase diagram. 

\subsection{Gaussian approximation and beyond the MF results}~\label{sub-sec: 3-mode cumulant expansion}
So far, we have studied the MF-phase diagram of the open system and investigated the disspation-stabilized phases in contrast to the closed system. Moreover, using Keldysh approach, we have been able to investigate the spectral signatures of each phase and delimit the onset of a PT, beyond the MF. 

As can be inferred from the Langevin EoM in Eq~\ref{eq:EoM 3-mode cavity} of Appendix~\ref{app:MF_EOM}, the quartic interaction leads to an infinite hierarchy of moments. Therefore, any semi-analytic or numerical calculations require a truncation of this hierarchy. The mean-field treatment ignores the correlation via the factorization approximation, hence truncating the cumulant expansions to the $1^\textrm{st}$-order. Recent studies however, show that the inclusion of quantum correlations lead to marked deviations from the mean-field results, especially close to the phase transition points~\cite{Reiter2020}. So far, there have been several approaches to include the effect of higher-order correlation including exact diagonalization, diagrammatic expansions, functional renormalization group, numerical or density matrix renormalization group analysis, and phase space methods.~\cite{Schliech2001,Bulla_2003,Schollwock_2005,Sieberer_2013,Aoki_2014,Tauber_2014,Mascarenhas_2015,Jin_2016,Mathey_2020,Arndt_2021}. 

In this section, we extend the EoM to include the moment dynamics up to the $2^\textrm{nd}$-order, while assuming a vanishing $3^\textrm{rd}$-order, to find the Gaussian approximation of the NESS with an emphasis on the robustness of the LC-phase. Further details for this specific problem and the comparison between the results of this approach with the MF and the full density matrix calculations for a single-mode Kerr cavity can be found in the Appendix~\ref{appendix:2nd_cumulant}. 

Figure~\ref{fig:3mode_2nd_cumulant}(a)-(c) shows the population of the side modes ($\braket{\hat{a}_{1,3}^\dagger \hat{a}_{1,3}}$ in blue), the pumped mode ($\braket{\hat{a}_2^\dagger \hat{a}_2}$ in red), and the correlation between the generated pairs in the side modes ($|\braket{\hat{a}_1 \hat{a}_3}|$ in brown), for $\Delta_2 = -5, 0, +5$, respectively.

\begin{figure}[h]
    \includegraphics{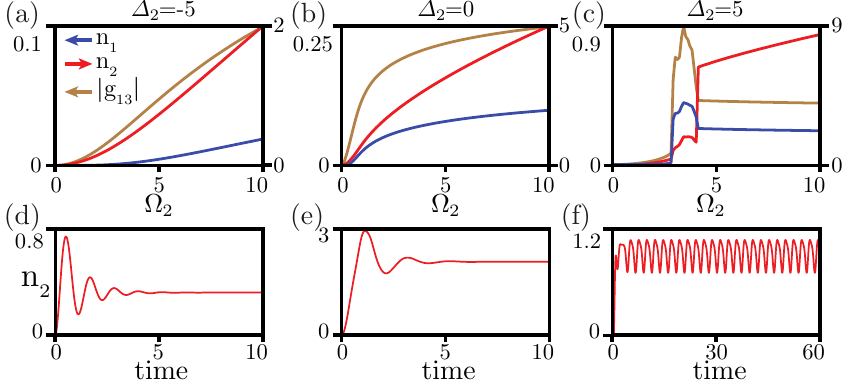}
    \centering
    \caption{(a)-(c) Population of the pumped mode ($\braket{\hat{a}_2^\dagger \hat{a}_2}$ in red), the un-pumped modes ($\braket{\hat{a}_1^\dagger \hat{a}_1}$ in blue), and the correlation between the side-modes ($|\braket{\hat{a}_1 \hat{a}_3}|$ in brown) as a function of the pumping rate $\Omega_2$ at detuning of $\Delta_2 = -5, 0, +5$, respectively. (d)-(c) The time trace of the pumped mode population for corresponding detuning and at fixed pump rate of $\Omega_2 = 3$.}
    \label{fig:3mode_2nd_cumulant}
\end{figure}

For panels (a),(b) corresponding to the MF uniform phase, \textit{i.e.} one solution for the pumped mode and no occupation of the side modes, the Gaussian approximation results are in good agreement with the MF ones depicted in Fig.~\ref{fig:3-mode MF}(a),(b).
For the side-modes however, the MF predicts zero population while the $2^\textrm{nd}$-cumulant suggests a finite but low value. This can be understood in terms of the quantum fluctuations that had been ignored in the MF, and partially resumed in the Gaussian approximation. Besides, the weak correlation between the side modes, \textit{i.e.} $|g_{13}|$, reinforces that interpretation.

For $\Delta_2 = +5$ however, the MF (Fig.~\ref{fig:3-mode MF}(c)) and the Gaussian approximation (Fig.~\ref{fig:3mode_2nd_cumulant}(c)) show marked differences. While the MF shows a multistable behavior, a $1^\textrm{st}$-order phase transitions, and an LC behavior for the side modes, the Gaussian approximation results are continuous for all modes. For the side modes and before the LC-phase, the trend is quite similar to the other detuning in panels (a) and (b), \textit{i.e.} a finite but low population. Unlike the MF case however, the transition to the LC-phase is not a $1^\textrm{st}$-order as suggested by the MF and instead it shows a large but continuous change of the order parameter, $\braket{\hat{a}_1^\dagger \hat{a}_1}$. This transition is accompanied with a maximum in the correlation between the side modes,\textit{i.e.} $\braket{\hat{a}_1 \hat{a}_3}$, a quantity that signifies the correlated photon-pair generation within this phase. It is interesting to note that this correlation is low before the LC-phase and drops but remains finite after the LC. Moreover, the MF predicts a uniform phase again after the LC, where there is no population in the side modes, while the Gaussian approximation delineates that a finite population after the LC is the correlation effect, only.

Figure~\ref{fig:3mode_2nd_cumulant}(d)-(f) presents the temporal evolution of the pumped mode for the corresponding detuning in (a)-(c) at a fixed pumping rate of $\Omega_2 = 3$ starting from vacuum, \textit{i.e.} $\braket{\hat{a}_m^\dagger \hat{a}_m} = 0$. For the uniform cases in (a),(b) the order parameter settles to a fixed value at the long-time limit after some transient behavior, a value which is very close to the corresponding MF value.

\begin{figure}[h]
    \includegraphics[width=\columnwidth]{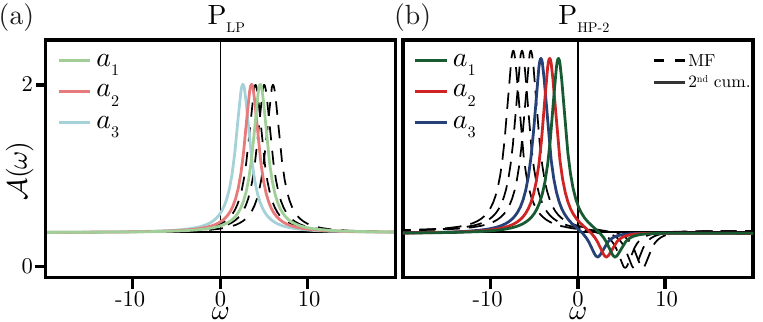}
    \centering
    \caption{(a),(b) The spectrum $\mathcal{A}(\omega)$ of the 3-mode cavity at $P_\textrm{LP}$ and $P_\textrm{HP}$ in Fig.~\ref{fig:phase diagram}. Using $2^\textrm{nd}$-cumulant approximation (solid lines) reveals a frequency pulling towards lower frequencies as compared to MF calculation (dashed lines). The effect is more pronounced in the high density phase where the corrections to the MF are bigger.}
    \label{fig:3mode_2nd_cumulant_specra}
\end{figure}

In contrary, the order parameter of the LC phase shows sustainable oscillations in time, after a short transient period, corroborating with the LC-phase features. It is interesting to note that, unlike the MF-predicted LC however, where the time-periodic behavior was only observable in the side-modes, here the coupled correlations lead to oscillatory behaviors in all correlations.

In Fig.~\ref{fig:3mode_2nd_cumulant_specra}, we compare the spectral functions of the modes obtained from the Keldysh approach based on the $2^\textrm{nd}$-cumulant results, with the ones determined from the MF-values. There is no qualitative change in the response and the peak swap between the LP and HP is still present. The $2^\textrm{nd}$-cumulant approximation reveals corrections to the eigen-frequencies of the system highlighted as a frequency pulling towards lower values. The effect is more pronounced in the HP, in agreement with the presence of stronger corrections to the MF results where the Gaussian approximation predicts a non-vanishing population for the un-pumped modes in contrary to zero MF-values.

\section{Keldysh Formalism }\label{sec:Keldysh}
The exact diagonalization of the Liouvillian super-operator suffers from the finite size effects, \textit{i.e.} the truncation of the Fock space, while the MF approach ignores quantum fluctuations. To include the correlation effects while approaching the thermodynamic limit (the typical validity range of a MF treatment), we employ Keldysh formalism~\cite{Sieberer2016}.

We readily write the rotated Keldysh action, associated with the Hamiltonian~\eqref{eq:3-mode Hamiltonian} and subjected to the dissipation~\eqref{eq:single-photon dissipator}, in the quantum and classical fields $ a_{m_{c,q}}$, for $m = 1,2,3$ as
\begin{equation}
    S_k(\vec{\alpha}_{c,q}) = \sum_{m=1}^3 S_0^m + S_\textrm{int} + S_\textrm{drive} + S_\gamma\,,
    \label{eq: keldysh action 3-mode rotated}
\end{equation}
where $S_0^m,S_\textrm{int},S_\textrm{drive},S_\gamma$ are defined in Appendix~\ref{app:S_rotated}. Here, we only highlight that the symmetry $a_1 \leftrightarrow a_3$ is still present in the action~\eqref{eq: keldysh action 3-mode rotated}.

We perform the saddle point approximation via $\partial S_k/ \partial \alpha_{c,q} = 0$, set the quantum fields to zero $\alpha_{m_q} = 0$, and obtain the EoM for the classical fields $\alpha_{c_m}$ [cf. Appendix~\ref{app:S_rotated}]. These equations coincide with the mean-field ones upon rotating back to physical fields, i.e., $\alpha_{m_c} = \sqrt{2}\alpha_m$.

To go beyond mean-field, we study the fluctuations around the MF stationary-states as $\alpha_m = \alpha_{m_c} + \delta\alpha_m$. We expand the Keldysh action in Eq.~\eqref{eq: keldysh action 3-mode rotated} and retain terms up to second order in fluctuations, \textit{i.e.} only the Gaussian parts. Therefore, the Gaussian action can be written in the normal form as
\begin{equation}
    S_G = \int_\omega \mathbf{\delta \Phi^\dagger(\omega)} M(\omega) \mathbf{\delta \Phi(\omega)},
\end{equation}
where we went to Fourier space with the 3-mode, 12 component, nambu-spinor $\mathbf{\delta \Phi(\omega)}$, and the matrix $M(\omega)$ is given by
\begin{equation}
    \label{eq:M_matrix_12x12}
    M_{12 \times 12} =  \frac{1}{2} \begin{pmatrix}
    [0]_{6\times 6} & [G^A(\omega)]_{6 \times 6}^{-1} \\ 
    [G^R(\omega)]_{6 \times 6}^{-1} & [P^K]_{6 \times 6}
    \end{pmatrix}\,,
\end{equation}
where $[G^A(\omega)]^{-1},[G^R(\omega)]^{-1},P^K(\omega)$ are the inverse of the advanced, retarded Green's functions, and the Keldysh component, respectively [cf. Appendix~\ref{app:6x6_GF}].

Using Gaussian integration we determine the single-mode action and the associated Green's functions. We here report solely the uniform phase case, $\alpha_2\ne 0,\,\alpha_{1,3}=0$ hence, the single-mode Green's functions has simple analytical expressions

\footnotesize
\begin{align}
    G_R^{2} & = -\frac{4(\omega + \Delta_2 + i\gamma_2 + U_0 n_2)}{4(\omega +\Delta_2 + i\gamma_2)(-\omega +\Delta_2 - i\gamma_2) + 8U_0\Delta_2n_2 + 3 U_0^2n_2^2} \label{eq:3mode_a2_GA}\\
    G_R^{m} & = \frac{4(\omega + \Delta_{\bar{m}} +i \gamma_{\bar{m}} + U_0 n_2)}{4(\omega + \Delta_{\bar{m}} + U_0 n_2 +i \gamma_{\bar{m}})(\omega - \Delta_{m} - U_0 n_2 +i \gamma_{m}) + U_0^2 n_2^2}\,,\label{eq:3mode_a13_GA}
\end{align}\normalsize
where $n_2= | \alpha_2|^2$ and $(m,{\bar{m}}) = (1,3),(3,1)$. 

From the poles of the Green's function we obtain the eigenvalues
\footnotesize
\begin{widetext}
\begin{align}
    \omega^{2} & = \pm\frac{1}{2} \sqrt{4\Delta_2^2 + 8 \Delta_2 U_0 n_2 + 3 U_0^2 n_2^2}+i \gamma_2 \\
    \omega^{m} & = \frac{\Delta_m-\Delta_{\bar{m}} +i \gamma_m+i \gamma_{\bar{m}}}{2} 
    \pm\frac{1}{2}\sqrt{(\Delta_m+\Delta_{\bar{m}}+i\gamma_m-i\gamma_{\bar{m}})^2 + 4 U_0 n_2 (\Delta_m+\Delta_{\bar{m}}+i \gamma_m-i \gamma_{\bar{m}}) + 3 U_0^2 n_2^2} \notag
\end{align}
\end{widetext}
\normalsize
where $\omega^i$ is associated with the $i^\textrm{th}$ cavity mode and $i=1,2,3$.

First, we notice that the pumped mode behaves as a single driven Kerr-oscillator whereas the symmetric modes $a_{1,3}$, even though empty and not pumped, present a non-trivial Green's function, and their Green's functions are ``coupled''. Second, the eigenvalues we obtained from the Green's functions coincide, upon the substitutions $\alpha_m = \sqrt{2}\alpha_m^{MF}$, $\gamma_0 = \gamma_1 = \gamma_2 = \gamma_3$, $\Delta_1 + \Delta_3 = 2(\Delta_2-\delta_D)$, with those obtained from the stability matrix [cf. Appendix~\ref{app:MF_EOM}].

\section{Experimental Realizations}


We consider a multi-mode spherical Fabry-Perot cavity, as sketched in Fig.~\ref{fig:schematics_intro}. Depending on the mirrors radii of curvature and the cavity length, the cavity leads to an effective harmonic trap for the photons and sets the bare cavity spectrum, \textit{i.e.} $\omega_{1,2,\cdots}$~\cite{siegman1986}. Due to the finite reflectively from the mirrors, there will be a cavity decay rate $\gamma_n$ corresponding to the $n^\textrm{th}$-mode, which can be approximated with a fixed rate $\gamma_0$, over a finite frequency range of our interest. The self- and cross-mode interactions are induced via a non-linear medium inside the cavity, ranging from ultracold atomic gas trapped along the cavity axis \cite{Koller2015,Vaidye2018,Guo2019}, to excitons~\cite{Carusotto2013} or highly-excited Rydberg excitons in solid state. By re-writing the dynamics in terms of the polaritons, \textit{i.e.} the hybrid particles of cavity modes and the material degrees of freedom, we can get the final dynamics of Eq.~\ref{eq:3-mode Hamiltonian}. In brief, $\hat{a}_n,\hat{a}_n^\dagger$  represent the annihilation and creation operator of a polariton in the $n^\textrm{th}$-cavity mode, and $U_0$ is the effective interaction between polaritons after approximating the two-body interaction, or the third-order optical non-linearity, with the short-range contact interaction.
By tuning the relative frequency between the \emph{atoms} the cavity modes, the interaction can can be attractive or repulsive. A laser with frequency $\omega_L$ coherently drives the $2^\textrm{nd}$-cavity mode. Note that due to the orthogonality of the Gauss-Lauguerre modes of spherical cavities, excitation of a specific mode is possible. Although here we only focused on a canonical cavity-QED setup, recent developments of the multi-mode superconducting cavities with parametrically-induced self- and cross-Kerr non-linearity allows ones to engineer similar Hamiltonian using the circuit-QED toolbox, as well~\cite{hung2021quantum}.

\section{Conclusion and Outlook}~\label{sec:conclusion}
For a long time it has been speculated if the limit cycle, a recurring phenomenon in many non-linear classical and quantum many-body systems, can be classified as a time crystal. Although the semi-classical MF treatment of several many-body systems suggests the appearance of periodic long-time behavior, however, the robustness of these periodic dynamics, against quantum fluctuations has been an open question.
In this work, we investigated this question in a canonical yet important system, \textit{i.e.} a multi-mode cavity with self and cross-Kerr non-linearity. 

We used the Keldysh formalism and extended the MF results via a higher-order cumulant expansion. Both of these methods confirmed the robustness of the LC phase phase and the overall topology of the phase diagram. Additionally, we revealed corrections to the MF results, \textit{e.g.} the change of the correlation spectra, the non-vanishing population of the side modes, and the oscillating cross-correlations.

The results and the methods presented in this work can be employed in the study of a versatile group of the many-body systems to understand the effects of higher order correlations and investigate the robustness of dissipation-stabilized MF phases, including the limit cycle.
It is straightforward to extend this work to study the dynamical behavior of such systems subject to a parametric drive with a modulated amplitude, and investigate the system transition from a stationary dynamics, to a limit cycle, to a chaotic phase as a function of the modulation frequency and depth.

Another interesting direction is to extend the system size either by including more modes in a single cavity or considering an array of coupled cavities with self- and cross-Kerr non-linearity. When subject to a two-photon parametric drive, instead of a single-photon as studied here, the $Z_2$ symmetry of the modes are preserved and new phase transitions could appear due to the spontaneous breaking of this symmetry~\cite{Bartolo2016, Puri2017}. It would be interesting to examine if for some parameter range this symmetry breaking can be accompanied with a limit-cycle phase. Since such a system, at certain limits, can approximate a spin Ising Hamiltonian, it is interesting to investigate the notion of the LC phase on the magnetization. Finally, one can explore the ultimate behavior of cascaded quantum systems where the limit cycle output of the first sub-system behaves as a parametric drive for the second one. This could be thought as an alternative approach for creating time crystals using discrete time-symmetric drive.
\section*{Acknowledgment}
MS would like to thank O. Zilberberg for fruitful discussions and acknowledges financial support from the Swiss National Science Foundation through the Grant No. PP00P2\_163818. HA acknowledges the financial supports from Baden-W\"urttemberg Stiftung Eliteprogram award and the Purdue University Startup fund. SFY would like to thank the NSF through the CUA PFC grant
PHY-1734011 and the AFOSR via FA9550-19-1-0233, and the NSF via PHY-1912607.

\bibliography{ref}

\clearpage
\appendix

\section{The mean-field EoM, dynamical stability, and covariance matrix}~\label{app:MF_EOM}
Directly from the master equation of Eq.~\ref{eq:Liouvillian} we obtain the Heisenberg-Langevin equations of motion for the field operators as
\begin{widetext}
\begin{align}~\label{eq:EoM 3-mode cavity}
     \frac{d}{dt}\hat{a}_1 & = -i\left(\Delta_1 - i\gamma_1 \right) \hat{a}_1 -i U_0 \left[\left(\hat{a}_1^\dagger \hat{a}_1 + 2 \hat{a}_2^\dagger \hat{a}_2 + 2 \hat{a}_3^\dagger \hat{a}_3 \right)\hat{a}_1  + \hat{a}_2^2 \hat{a}_3^\dagger \right] + \sqrt{2\gamma_1} ~ \hat{\xi}_1(t)\,,  \\
      \frac{d}{dt} \hat{a}_2 & = -i\left(\Delta_2  -i\gamma_2 \right) \hat{a}_2 -i U_0 \left[ \left(\hat{a}_2^\dagger \hat{a}_2  + 2 \hat{a}_1^\dagger \hat{a}_1 +  2 \hat{a}_3^\dagger \hat{a}_3 \right)\hat{a}_2 + 2 \hat{a}_2^\dagger \hat{a}_1 \hat{a}_3 \right] -i \Omega_2 + \sqrt{2\gamma_2} ~ \hat{\xi}_2(t) \,, \notag \\
      \frac{d}{dt}\hat{a}_3 & = -i \left(\Delta_3 - i\gamma_3 \right) \hat{a}_3 -i U_0 \left[\left( \hat{a}_3^\dagger \hat{a}_3 + 2 \hat{a}_2^\dagger \hat{a}_2 + 2 \hat{a}_1^\dagger \hat{a}_1 \right) \hat{a}_3  + \hat{a}_2^2 \hat{a}_1^\dagger  \right] + \sqrt{2\gamma_3} ~ \hat{\xi}_3(t) \,, \notag
\end{align}
\end{widetext}
where $\hat{\xi}_m(t)$ is a zero-mean noise operator with the correlations of $\braket{\hat{\xi}_m(t_1) \hat{\xi}_n^\dagger (t_2)} = \delta_{mn} ~ \delta(t_1 - t_2)$ and $\braket{\hat{\xi}_m^\dagger (t_1) \hat{\xi}_n(t_2)} = 0$.

Ignoring the quantum correlations in the aforementioned EoM and employing the factorization assumption, one can get a set of coupled non-linear equations for describing the MF of $\braket{\hat{a}_m} = \alpha_m e ^{i\phi_m}$ as
\begin{widetext}
\begin{align}~\label{eq:MF 3-mode cavity}
     \frac{d}{dt}\braket{\hat{a}_{1}} & = -i\left(\Delta_1 - i\gamma_1 \right) \braket{\hat{a}_1} -i U_0 \left[ \left(|\braket{\hat{a}_1}|^2 + 2 |\braket{\hat{a}_2}|^2 + 2 |\braket{\hat{a}_3}|^2 \right) \braket{\hat{a}_1}  + \braket{\hat{a}_2}^2 \braket{\hat{a}_3}^* \right] \,,  \\
      \frac{d}{dt}\braket{\hat{a}_2} & = -i \left(\Delta_2  -i\gamma_2 \right) \braket{\hat{a}_2} -i U_0 \left[ \left(|\braket{\hat{a}_2}|^2  + 2|\braket{\hat{a}_1}|^2 +  2|\braket{\hat{a}_3}|^2 \right)\braket{\hat{a}_2} + 2 \braket{\hat{a}_2}^* \braket{\hat{a}_1} \braket{\hat{a}_3}  \right] -i \Omega_2 \,, \notag \\
      \frac{d}{dt} \braket{\hat{a}_3} & = -i \left(\Delta_3 - i\gamma_3 \right) \braket{\hat{a}_3} -i U_0 \left[\left( |\braket{\hat{a}_3}|^2 + 2 |\braket{\hat{a}_2}|^2 + 2 |\braket{\hat{a}_1}|^2 \right) \braket{\hat{a}_3} + \braket{\hat{a}_2}^2 \braket{\hat{a}_1}^*  \right] \,. \notag
\end{align}
\end{widetext}
As can be seen, while the coherent drive restricts ($2\phi_2 - \phi_1 - \phi_3$), hence breaking the U(1)-symmetry of this field, there is no additional constraint on the phase of the un-pumped modes. This phase freedom leads to the emergence of the LC-phase. 
The stagnation points of the aforementioned EoM is determined as LHS = 0. 
When dynamics are contractive at a particular stagnation point, field operators can be linearized around that MF with a fluctuation vector $[\delta \hat{\Phi}]$ as 
\begin{equation}~\label{eq:fluctuation vector}
    [\delta \hat{\Phi}] = \begin{bmatrix}
    \delta \hat{a}_1 & 
    \delta \hat{a}_2 & 
    \delta \hat{a}_3 &
    \delta \hat{a}_1^\dagger &
    \delta \hat{a}_2^\dagger & 
    \delta \hat{a}_3^\dagger
    \end{bmatrix}^T \,,
\end{equation}
where the superscript $T$ means the matrix transpose.

To evaluate the stability of these solutions we employ the dynamical stability analysis. 
$\mathcal{M}$, \textit{i.e.} the Bogoliubov matrix of the small excitation, has the following structure
\begin{equation}~\label{eq:Bogoliuov Matrix}
    \mathcal{M} = \begin{bmatrix}
    R & S \\
    S^* & R^*
    \end{bmatrix} \,,
\end{equation}
where 
\begin{widetext}~\label{eq:R-matrix}
\begin{align*}
   R = -i \begin{bmatrix}
    (\Delta_1 - i\gamma_1) + 2 U_0 \mathcal{N} & 2 U_0 \left(\braket{a_1} \braket{a_2}^* + \braket{a_2} \braket{a_3}^* \right) & 2 U_0 \braket{a_1} \braket{a_3}^* \\
    2U_0 \left(\braket{a_1}^* \braket{a_2} + \braket{a_2}^* \braket{a_3} \right) & (\Delta_2 - i\gamma_2) + 2U_0 \mathcal{N} & 2U_0 \left(\braket{a_1} \braket{a_2}^* + \braket{a_2} \braket{a_3}^* \right) \\
    2U_0 \braket{a_1}^* \braket{a_3} & 2U_0 \left(\braket{a_1}^* \braket{a_2} + \braket{a_2}^* \braket{a_3} \right) & \left(\Delta_3 - i\gamma_3 \right) + 2U_0 \mathcal{N}
    \end{bmatrix}, 
\end{align*}
\end{widetext}
and 

\begin{widetext}~\label{eq:S-matrix}
 \begin{align}
    S = -i U_0
    \begin{bmatrix}
    \braket{a_1}^2 & 2\braket{a_1} \braket{a_2} & \left(\braket{a_2}^2 + 2 \braket{a_1} \braket{a_3}\right) \\
    2\braket{a_1} \braket{a_2} & \left(\braket{a_2}^2 + 2\braket{a_1} \braket{a_3} \right) & 2 \braket{a_2} \braket{a_3} \\
    \left(\braket{a_2}^2 + 2 \braket{a_1} \braket{a_3} \right) & 2 \braket{a_2} \braket{a_3} & \braket{a_3}^2
    \end{bmatrix}\, ,
   \end{align}
\end{widetext}
for $\mathcal{N} = \alpha_{1}^2 + \alpha_{2}^2 + \alpha_{3}^2$ being the mean-value of the total number of the photons in the cavity. 

When MF is dynamically-stable the Bogoliubov matrix $\mathcal{M}$ is negative-definite.
For the stationary-state of the stable solutions we can define the covariance matrix as $\Gamma_a(\omega) = \braket{[\delta \hat{\Phi}(\omega)] [\delta\hat{\Phi}(\omega)]^\dagger}$, with the following entries
\begin{equation}~\label{eq:correlation matrix from Bog.}
    \Gamma_{a_{mn}}(\omega) = \mathcal{F}\{\lim_{t \longrightarrow \infty} \braket{\delta \hat{\Phi}_m(t+\tau) \delta \hat{\Phi}_n^\dagger(t)}\}_\tau \,.
\end{equation}
From this equation it is clear that the spectrum of the correlation matrix is closely related to the noise operator spectrum. The diagonal entries of the covariance matrix are related to the cavity transmission spectrum (auto-correlations), and the off-diagonal entries signify the cross-correlations between the modes hence, are related to the entanglement between the modes.

\subsection{uniform phase}
Within this phase, only the $2^\textrm{nd}$-mode has a non-zero MF and $\alpha_{1,3} = 0$, making $R,S$ diagonal and anti-diagonal matrices, respectively. The eigenvalues of the dynamical stability matrix $\mathcal{M}$ for identical loss rates of the cavity modes $\gamma_0$, are 
\begin{widetext}~\label{eq:eigenvlaue uniform phase}
 \begin{align}
    \lambda_{1,2} & = -\gamma_0 \pm i \sqrt{3 U_0^2 n_2^2 + 4 U_0 \Delta_2 n_2 + \Delta_2^2} \, , \\ \notag
    \lambda_{3,4,5,6} & = -\gamma_0 \pm i \left( \frac{\Delta_1 - \Delta_3}{2} \pm  \sqrt{3 U_0^2 n_2^2 + 4 U_0 (\Delta_2 - \delta_D) n_2 + (\Delta_2 - \delta_D)^2} \right)\, ,
    \end{align}
\end{widetext}
where the dispersion parameter $\delta_D$ is defined as $2\delta_D = 2\Delta_2 - (\Delta_1 + \Delta_3) = 2\omega_2 - \omega_1 -\omega_3$.

As can be seen, for $U_0 \Delta_2 \ge 0$ and $U_0 (\Delta_2 - \delta_D) \ge 0$, matrix $\mathcal{M}$ is negative definite in the uniform phase hence, stagnation points are dynamically \emph{stable}.

The many-body system is within the uniform phase when $\Omega_2$ is either small or large. If the former, the cross-interaction compared to the self-interaction is so small that the pair generation process cannot start. In the latter, the number of particles in the pumped mode is so large that the self-interaction shifts the pumped mode out of resonance by several $\gamma_0$ such that the inter-modal scattering ceases. In other words, in the extreme of a strong pump, the large self-interaction dominates all many-body interactions hence, pushing the system to the single-body dynamics again.

On the single-mode branch only, one gets
\begin{equation}~\label{eq:uniform-phase slope}
    \frac{dn_2}{d\Omega_2} = \frac{2\Omega_2}{3 U_0^2 n_2^2 + 4 U_0 \Delta_2 n_2 + \left(\Delta_2^2 + \gamma_2^2 \right)  } \, .
\end{equation}

When $U_0 \Delta_2 \le 0$, there might be points that the slope first diverges and later changes the sign. 
These turning points in the MF dynamics signify the existence of multiple MF attractors. More specifically, the boundaries of the uniform phase, can be determined as
\begin{equation}~\label{eq:turning points}
    n_2 = \frac{- 2 U_0 \Delta_2 \mp \sqrt{U_0^2 \left(\Delta_2^2 - 3 \gamma_2^2 \right)}}{3U_0^2} \, .
\end{equation}
The above equation determines that for $\delta_D = 0$ the multi-stability exists when $\Delta_2 \ge \gamma_2 \sqrt{3}$.

A similar argument for a dispersive cavity, \textit{i.e.} $\delta_D \ne 0$, shows that the multi-stability can exist for $U_0 (\Delta_2 - \delta_D) \le 0$ and $|\Delta_2 - \delta_D| \ge \gamma_2 \sqrt{3}$.

The covariance matrix within this uniform phase has the following general form 
\begin{widetext}~\label{eq:covariance uniform-phase}
\begin{align*}
   \Gamma_a(\omega) = \begin{bmatrix}
   \braket{\hat{a}_1(\omega) \hat{a}_1^\dagger(-\omega)} & 0 & 0 & 0 & 0 & \braket{\hat{a}_1(\omega) \hat{a}_3(-\omega)} \\ \notag
   0 & \braket{\hat{a}_2(\omega) \hat{a}_2^\dagger(-\omega)} & 0 & 0 & \braket{\hat{a}_2(\omega) \hat{a}_2(-\omega)} & 0 \\ \notag
   0 & 0 & \braket{\hat{a}_3(\omega) \hat{a}_3^\dagger(-\omega)} & \braket{\hat{a}_3(\omega) \hat{a}_1(-\omega)} & 0 & 0 \\ \notag
   0 & 0 & \braket{\hat{a}_1^\dagger(\omega) \hat{a}_3^\dagger(-\omega)} &
   \braket{\hat{a}_1^\dagger(\omega) \hat{a}_1(-\omega)} & 0 & 0 \\ \notag
   0 & \braket{\hat{a}_2^\dagger(\omega) \hat{a}_2^\dagger(-\omega)} & 0 & 0 & \braket{\hat{a}_2^\dagger(\omega) \hat{a}_2(-\omega)} & 0 \\ \notag
   \braket{\hat{a}_3^\dagger(\omega) \hat{a}_1^\dagger(-\omega)} & 0 & 0 & 0 & 0 & \braket{\hat{a}_3^\dagger(\omega) \hat{a}_3(-\omega)}
   \end{bmatrix} \, .
\end{align*}
\end{widetext}
The structure of this matrix implies that the $2^\textrm{nd}$-mode has no correlation with the side modes. That physically is consistent with the picture of these un-pumped modes not being populated through the parametric process via the $2^\textrm{nd}$-mode.

Consequently, the covariance matrix for the quadratures gets the following block diagonal form
\begin{equation}
    \Gamma_\textrm{sym}(\omega) = \begin{bmatrix}
    \Gamma_2(\omega) & \textbf{0}_{4 \times 4} \\
    \textbf{0}_{2 \times 2} & \Gamma_\pm(\omega) 
    \end{bmatrix} \, .
\end{equation}
The above form emphasizes again that the pumped mode has no correlation with the other two side modes.
\subsection{Limit-Cycle phase}
As discussed before, while the U(1)-symmetry of the un-driven system is broken by a coherent drive, the un-pumped modes still have some phase freedom remained since $ \Phi_0 = 2\phi_2 - \phi_1 - \phi_3$, is the only constraint imposed by the coherent pump.

Within this phase there is no stationary state and the long-time limit of the $1^\textrm{st},3^\textrm{rd}$ MF has an oscillatory behavior as $e^{\pm i t \omega_\textrm{LC}}$. In other words, due to the aforementioned phase constraint, if $\hat{a}_1$ oscillates as $e^{i t \omega_\textrm{LC}}$, $\hat{a}_3$ should vary as $e^{-i t \omega_\textrm{LC}}$. Going back to the Eq.~\ref{eq:EoM 3-mode cavity}, one can see that this oscillation can be interpreted in terms of a re-normalized detuning of the parametrically-populated modes as $\tilde{\Delta}_{1,3} = \Delta_{1,3} \pm \omega_\textrm{LC}$.

\section{Rotated Keldysh action}
\label{app:S_rotated}
In this appendix, we give the explicit expressions for the rotated Keldysh actions used in the Eq.~\eqref{eq: keldysh action 3-mode rotated} of the main text.

The Keldysh action of the $m^\textrm{th}$-mode with the self-interaction reads as
\begin{widetext}
\begin{align}
S_0^m = \alpha_{m_c}^* \left(i \frac{\partial}{\partial t} - \Delta_m \right) \alpha_{m_q}
+ \alpha_{m_q}^* \left(i \frac{\partial}{\partial t} - \Delta_m \right) \alpha_{m_c}
- \frac{U_0}{2} \left(|\alpha_{m_c}|^2 + |\alpha_{m_q}|^2 \right)\left(\alpha_{m_c} \alpha_{m_q}^* + \alpha_{m_c}^* \alpha_{m_q} \right) \, .
\end{align}
\end{widetext}
The action due to the cross-term interactions between the modes reads as
\begin{widetext}
\begin{align}
S_\textrm{int}   = & -U_0 \left[\left(|\alpha_{1_c}|^2 + |\alpha_{1_q}|^2\right)\left(\alpha_{2_c} \alpha_{2_q}^*  + \alpha_{2_c}^* \alpha_{2_q} + \alpha_{3_c} \alpha_{3_q}^* + \alpha_{3_c}^* \alpha_{3_q}\right) \right] \\ \notag
& -U_0 \left[\left(|\alpha_{3_c}|^2 + |\alpha_{3_q}|^2\right)\left(\alpha_{2_c} \alpha_{2_q}^*  + \alpha_{2_c}^* \alpha_{2_q} + \alpha_{1_c} \alpha_{1_q}^* + \alpha_{1_c}^* \alpha_{1_q}\right) \right]  \notag \\
& -U_0 \left[\left(|\alpha_{2_c}|^2 + |\alpha_{2_q}|^2\right)\left(\alpha_{1_c} \alpha_{1_q}^*  + \alpha_{1_c}^* \alpha_{1_q} + \alpha_{3_c} \alpha_{3_q}^* + \alpha_{3_c}^* \alpha_{3_q}\right)\right]  \notag \\
& - \frac{U_0}{2} \left[\left(\alpha_{2_c}^{*^2} + \alpha_{2_q}^{*^2} \right) \left(\alpha_{1_c} \alpha_{3_q} + \alpha_{1_q} \alpha_{3_c} \right) + 2 \alpha_{2_c}^* \alpha_{2_q}^* \left(\alpha_{1_c} \alpha_{3_c} + \alpha_{1_q} \alpha_{3_q} \right)   \right]  + c.c. \, .
    %
    %
\end{align}
\end{widetext}
The coherent drive action is 
\begin{equation}
S_{drive} = - \Omega_2 \sqrt{2} \left(\alpha_{2_q} + \alpha_{2_q}^* \right) \, .
\end{equation}
And finally the action due to the Lindblad dissipator reads as
\begin{equation}
S_\gamma  = i \sum_{m=1}^3 \gamma_m \left(2|\alpha_{m_q}|^2 + \alpha_{m_c} \alpha_{m_q}^* - \alpha_{m_c}^* \alpha_{m_q} \right). 
\end{equation}  
The action is stationary at the saddle point, determined via $\partial S_k/ \partial \alpha_{c,q} = 0$, which leads to $\alpha_{m_q} = 0$ and the following equations of motion for $\alpha_{m_c}$
\begin{widetext}~\label{eq:Kelysh saddle point}
\begin{align}
     \frac{d}{dt}\alpha_{1_c} & = -i\left(\Delta_1 - i\gamma_1 \right) \alpha_{1_c} -i \frac{U_0}{2} \left[\left(|\alpha_{1_c}|^2 + 2 |\alpha_{2_c}|^2 + 2 |\alpha_{3_c}|^2 \right)\alpha_{1_c}  + \alpha_{2_c}^2 \alpha_{3_c}^* \right] \,,  \\
      \frac{d}{dt}\alpha_{2_c} & = -i \left(\Delta_2  -i\gamma_2 \right) \alpha_{2_c} -i \frac{U_0}{2} \left[ \left(|\alpha_{2_c}|^2  + 2|\alpha_{1_c}|^2 +  2|\alpha_{3_c}|^2 \right)\alpha_{2} + 2 \alpha_{2_c}^* \alpha_{1_c} \alpha_{3_c}  \right] -i \Omega_2 \sqrt{2} \,, \notag \\
      \frac{d}{dt}\alpha_{3_c} & = -i \left(\Delta_3 - i\gamma_3 \right) \alpha_{3_c} -i \frac{U_0}{2} \left[\left( |\alpha_{3_c}|^2 + 2 |\alpha_{2_c}|^2 + 2 |\alpha_{1_c}|^2 \right) \alpha_{3_c} + \alpha_{2_c}^2 \alpha_{1_c}^*  \right] \,. \notag
\end{align}
\end{widetext}
Comparing the EoM for $\alpha_{m_c}$ of the action saddle points with Eq.~\ref{eq:EoM 3-mode cavity} for the MFs, one can see that $\alpha_{m_c} = \sqrt{2} \braket{\hat{a}_m}$, which further clarifies the meaning of the classical fields in terms of the mean-fields.

\subsection{Approximated Gaussian Action}
\label{app:fluc_action}
As mentioned in the text, the effects of quantum fluctuations $\delta \alpha_{m_{c,q}}$, can be included by expending the Keldysh action around the saddle points. Since the Hamiltonian of Eq.~\ref{eq:3-mode Hamiltonian} is quartic, the expansion has terms linear - quartic in fluctuations, in general. 
Here, we report the explicit form of the Keldysh action up to second order in fluctuations, \textit{i.e.} only the Gaussian parts
\begin{widetext}
\begin{align}
S_k^{(2)} & = \sum_{m=1}^3 \mathbf{\delta \alpha_{m_c}^*} \left[\left(i \frac{\partial}{\partial t} - \Delta_m\right)  - U_0 \left( |\alpha_{1_c}|^2 + |\alpha_{2_c}|^2 + |\alpha_{3_c}|^2\right) - i\gamma_m \right] \mathbf{\delta \alpha_{m_q}} + i2\gamma_m \mathbf{\delta \alpha_{m_q}^*} \mathbf{\delta \alpha_{m_q}} \\ \notag
& + \sum_{m=1}^3 \mathbf{\delta \alpha_{m_q}^*} \left[\left(i \frac{\partial}{\partial t} - \Delta_m\right)  - U_0 \left( |\alpha_{1_c}|^2 + |\alpha_{2_c}|^2 + |\alpha_{3_c}|^2\right) + i\gamma_m \right] \mathbf{\delta \alpha_{m_c}} \\ \nonumber
& - U_0 \left( \alpha_{1_c} \alpha_{2_c}^* + \alpha_{2_c} \alpha_{3_c}^* \right) \left(\mathbf{\delta \alpha_{1_c}^*} \mathbf{\delta \alpha_{2_q}} + \mathbf{\delta \alpha_{2_c}} \mathbf{\delta \alpha_{1_q}^*} + \mathbf{\delta \alpha_{2_c}^*} \mathbf{\delta \alpha_{3_q}} + \mathbf{\delta \alpha_{3_c}} \mathbf{\delta \alpha_{2_q}^*} \right)  
%
%
%
%
- U_0 ~ \alpha_{1_c} \alpha_{2_c} \left(\mathbf{\delta \alpha_{1_c}^*} \mathbf{\delta \alpha_{2_q}^*} + \mathbf{\delta \alpha_{2_c}^*} \mathbf{\delta \alpha_{1_q}^*}  \right) + c.c. \\ \notag
%
%
%
& - U_0 ~\alpha_{2_c} \alpha_{3_c} \left(\mathbf{\delta \alpha_{2_c}^*} \mathbf{\delta \alpha_{3_q}^*} + \mathbf{\delta \alpha_{3_c}^*} \mathbf{\delta \alpha_{2_q}^*}  \right) 
%
%
%
- U_0 ~ \alpha_{1_c} \alpha_{3_c}^* \left(\mathbf{\delta \alpha_{1_c}^*} \mathbf{\delta \alpha_{3_q}} + \mathbf{\delta \alpha_{3_c}} \mathbf{\delta \alpha_{1_q}^*}  \right) + c.c. \\ \notag
%
%
%
& - \frac{U_0}{2} \left(\alpha_{1_c}^2 \mathbf{\delta \alpha_{1_c}^*} \mathbf{\delta \alpha_{1_q}^*} + \alpha_{1_c}^{*^2} \mathbf{\delta \alpha_{1_c}} \mathbf{\delta \alpha_{1_q}} \right)
- \frac{U_0}{2} \left(\alpha_{3_c}^2 \mathbf{\delta \alpha_{3_c}^*} \mathbf{\delta \alpha_{3_q}^*} + \alpha_{3_c}^{*^2} \mathbf{\delta \alpha_{3_c}} \mathbf{\delta \alpha_{3_q}} \right) \\ \nonumber
& - \frac{U_0}{2} \left(\alpha_{2_c}^{*^2} + 2 \alpha_{1_c}^* \alpha_{3_c}^* \right) \left(\mathbf{\delta \alpha_{1_c}} \mathbf{\delta \alpha_{3_q}} + \mathbf{\delta \alpha_{3_c}} \mathbf{\delta \alpha_{1_q}} + \mathbf{\delta \alpha_{2_c}} \mathbf{\delta \alpha_{2_q}} \right) + c.c. \, 
%
    \end{align}
\end{widetext}

\subsection{$M$ matrix coefficients}
\label{app:6x6_GF}

\begin{widetext}
In this appendix, we give the explicit form of the entries of the matrix $M$ in Eq.~\eqref{eq:M_matrix_12x12}

$[G^A(\omega)]^{-1}$ reads as

\noindent  \hrulefill 
\[
\begin{smallmatrix}
    \left(\omega - \Delta_1 - U_0 N - i\gamma_1\right) & - \frac{U_0}{2} \alpha_{1_c}^2 & - U_0 \left(\alpha_{1_c} \alpha_{2_c}^* + \alpha_{2_c} \alpha_{3_c}^* \right) & - U_0 \alpha_{1_c} \alpha_{2_c} & -U_0 \alpha_{1_c} \alpha_{3_c}^* & - \frac{U_0}{2} \left(\alpha_{2_c}^2 + 2 \alpha_{1_c} \alpha_{3_c}  \right) \\
    - \frac{U_0}{2} \alpha_{1_c}^{*^2} & \left(-\omega - \Delta_1 - U_0N + i\gamma_1 \right) & -U_0 \alpha_{1_c}^* \alpha_{2_c}^* & - U_0 \left(\alpha_{1_c}^* \alpha_{2_c} + \alpha_{2_c}^* \alpha_{3_c} \right) & - \frac{U_0}{2} \left(\alpha_{2_c}^{*^2} + 2 \alpha_{1_c}^* \alpha_{3_c}^* \right) & - U_0 \alpha_{1_c}^* \alpha_{3_c} \\
    - U_0 \left(\alpha_{1_c}^* \alpha_{2_c} + \alpha_{2_c}^* \alpha_{3_c} \right) & -U_0 \alpha_{1_c} \alpha_{2_c} & \left(\omega - \Delta_2 - U_0N - i\gamma_2 \right) & - \frac{U_0}{2} \left(\alpha_{2_c}^2 + 2 \alpha_{1_c} \alpha_{3_c} \right) & -U_0 \left(\alpha_{1_c} \alpha_{2_c}^* + \alpha_{2_c} \alpha_{3_c}^* \right) & - U_0 \alpha_{2_c} \alpha_{3_c} \\
    - U_0 \alpha_{1_c}^* \alpha_{2_c}^* & - U_0 \left(\alpha_{1_c} \alpha_{2_c}^* + \alpha_{2_c} \alpha_{3_c}^* \right) & - \frac{U_0}{2} \left(\alpha_{2_c}^{*^2} + 2\alpha_{1_c}^* \alpha_{3_c}^* \right) & \left(-\omega - \Delta_2 - U_0 N + i\gamma_2 \right) & -U_0 \alpha_{2_c}^* \alpha_{3_c}^* & - U_0 \left(\alpha_{1_c}^* \alpha_{2_c} + \alpha_{2_c}^* \alpha_{3_c} \right) \\
    - U_0 \alpha_{1_c}^* \alpha_{3_c} & -\frac{U_0}{2} \left(\alpha_{2_c}^2 + 2 \alpha_{1_c} \alpha_{3_c} \right) & - U_0 \left(\alpha_{1_c}^* \alpha_{2_c} + \alpha_{2_c}^* \alpha_{3_c} \right) & - U_0 \alpha_{2_c} \alpha_{3_c} & \left(\omega - \Delta_3 - U_0 N -i\gamma_3 \right) & -\frac{U_0}{2} \alpha_{3_c}^2 \\
    - \frac{U_0}{2}\left(\alpha_{2_c}^{*^2} + 2\alpha_{1_c}^* \alpha_{3_c}^* \right) & - U_0 \alpha_{1_c} \alpha_{3_c}^* & -U_0 \alpha_{2_c}^* \alpha_{3_c}^* & - U_0 \left(\alpha_{1_c} \alpha_{2_c}^* + \alpha_{2_c} \alpha_{3_c}^* \right) & - \frac{U_0}{2} \alpha_{3_c}^{*^2} & \left(-\omega - \Delta_3 - U_0 N + i\gamma_3 \right)
    \end{smallmatrix}
\]
\noindent \hrulefill 

$[G^R(\omega)]^{-1}$ reads as

\noindent  \hrulefill 
\[
\begin{smallmatrix}
\left(\omega - \Delta_1 - U_0 N + i\gamma_1 \right) & -\frac{U_0}{2} \alpha_{1_c}^2 & - U_0 \left(\alpha_{1_c} \alpha_{2_c}^* + \alpha_{2_c} \alpha_{3_c}^* \right) & - U_0 \alpha_{1_c} \alpha_{2_c} & - U_0 \alpha_{1_c} \alpha_{3_c}^* & -\frac{U_0}{2} \left(\alpha_{2_c}^2 + 2 \alpha_{1_c} \alpha_{3_c} \right) \\
- \frac{U_0}{2} \alpha_{1_c}^{*^2} & \left(- \omega - \Delta_1 - U_0 N - i\gamma_1 \right) & - U_0 \alpha_{1_c}^* \alpha_{2_c}^* & - U_0 \left(\alpha_{1_c}^* \alpha_{2_c} + \alpha_{2_c}^* \alpha_{3_c} \right) & -\frac{U_0}{2} \left(\alpha_{2_c}^{*^2} + 2 \alpha_{1_c}^* \alpha_{3_c}^* \right) &  - U_0 \alpha_{1_c}^* \alpha_{3_c} \\
- U_0 \left(\alpha_{1_c}^* \alpha_{2_c} + \alpha_{2_c}^* \alpha_{3_c} \right) & - U_0 \alpha_{1_c} \alpha_{2_c} & \left(\omega - \Delta_2 U_0 N + i\gamma_2 \right) & - \frac{U_0}{2} \left(\alpha_{2_c}^2 + 2\alpha_{1_c} \alpha_{3_c} \right) & - U_0 \left(\alpha_{1_c} \alpha_{2_c}^* + \alpha_{2_c} \alpha_{3_c}^* \right) & - U_0 \alpha_{2_c} \alpha_{3_c} \\
- U_0 \alpha_{1_c}^* \alpha_{2_c}^* & - U_0 \left(\alpha_{1_c} \alpha_{2_c}^* + \alpha_{2_c} \alpha_{3_c}^* \right) & - \frac{U_0}{2} \left(\alpha_{2_c}^{*^2}  + 2\alpha_{1_c}^* \alpha_{3_c}^* \right) & \left(-\omega - \Delta_2 - U_0 N - i\gamma_2 \right) & - U_0 \alpha_{2_c}^* \alpha_{3_c}^* & - U_0 \left(\alpha_{1_c}^* \alpha_{2_c} + \alpha_{2_c}^* \alpha_{3_c} \right) \\
- U_0 \alpha_{1_c}^* \alpha_{3_c} & - \frac{U_0}{2} \left(\alpha_{2_c}^2 + 2 \alpha_{1_c} \alpha_{3_c} \right) & - U_0 \left(\alpha_{1_c}^* \alpha_{2_c} + \alpha_{2_c}^* \alpha_{3_c} \right) & - U_0 \alpha_{2_c} \alpha_{3_c} & \left(\omega - \Delta_3 - U_0 N + i\gamma_3 \right) & - \frac{U_0}{2} \alpha_{3_c}^2 \\
- \frac{U_0}{2} \left(\alpha_{2_c}^{*^2} + 2\alpha_{1_c}^* \alpha_{3_c}^* \right) & - U_0 \alpha_{1_c} \alpha_{3_c}^* & - U_0 \alpha_{2_c}^* \alpha_{3_c}^* & - U_0 \left(\alpha_{1_c} \alpha_{2_c}^* + \alpha_{2_c} \alpha_{3_c}^*\right) & - \frac{U_0}{2} \alpha_{3_c}^{*^2} & \left(-\omega - \Delta_3 - U_0 N - i\gamma_3 \right)
\end{smallmatrix},
\]
\noindent \hrulefill 
\end{widetext}
where in the above equations $N = |\alpha_{1_c}|^2 + |\alpha_{2_c}|^2 + |\alpha_{3_c}|^2$ is the MF total number of particles in the cavity.

And finally we have $P^K$ as
\begin{equation}
    \begin{pmatrix}
    i2\gamma_1 [I]_{2\times 2} & 0 & 0 \\
    0 & i2\gamma_2 [I]_{2\times 2} & 0 \\
    0 & 0 & i2\gamma_3 [I]_{2\times 2}
    \end{pmatrix},
\end{equation}
for $[I]_{2 \times 2}$ being an identity matrix so rank 2.

\subsection{Single mode action}

We can determine the single-mode actions of each mode by Gaussian integration. Considering the steady state with empty mode $a_{1,3}$, Figs.~\ref{fig:3-mode MF}(a)-(b), simplifies the analytical expressions for the single mode Green's function to
\begin{widetext}
\begin{align}
    G_A^{2} & = -\frac{4(\omega + \Delta_2 - i\gamma_2 + U_0 \left| \alpha_2\right|^2)}{4(\omega +\Delta_2 - i\gamma_2)(-\omega +\Delta_2 + i\gamma_2) + 8U_0\Delta_2\left| \alpha_2\right|^2 + 3 U_0^2\left| \alpha_2\right|^4} \\
    G_A^{m} & = \frac{4(\omega + \Delta_{\bar{m}} -i \gamma_{\bar{m}} + U_0 \left| \alpha_2\right|^2)}{4(\omega + \Delta_{\bar{m}} + U_0 \left| \alpha_2\right|^2 -i \gamma_{\bar{m}})(\omega - \Delta_{m} - U_0 \left| \alpha_2\right|^2 -i \gamma_{m}) + U_0^2 \left| \alpha_2\right|^4}\,,
\end{align}
where  $(m,{\bar{m}}) = (1,3),(3,1)$. From the poles of the Green's function we obtain the eigenvalues
\begin{align}
    \epsilon^{2} & = \pm\frac{1}{2} \sqrt{4\Delta_2^2 + 8 \Delta_2 U_0 \left| \alpha_2\right|^2 + 3 U_0^2 \left| \alpha_2\right|^4}+i \gamma_2 \\
    \epsilon^{m} & = \frac{1}{2} (\pm\sqrt{(\Delta_m+\Delta_{\bar{m}}+i\gamma_m-i\gamma_{\bar{m}})^2 + 4 U_0 \left| \alpha_2\right|^2 (\Delta_m+\Delta_{\bar{m}}+i \gamma_m-i \gamma_{\bar{m}}) + 3 U_0^2 \left| \alpha_2\right|^4} + \Delta_m-\Delta_{\bar{m}} +i \gamma_m+i \gamma_{\bar{m}})
\end{align}
\end{widetext}
Interestingly, the pumped mode behaves as a single driven Kerr-oscillator whereas the symmetric modes $a_{1,3}$, even though empty, present a non-trivial Green's function.

\subsection{Spectral function}
\label{sub-sec:spectra}

\begin{figure}[h]
    \includegraphics[width=\columnwidth]{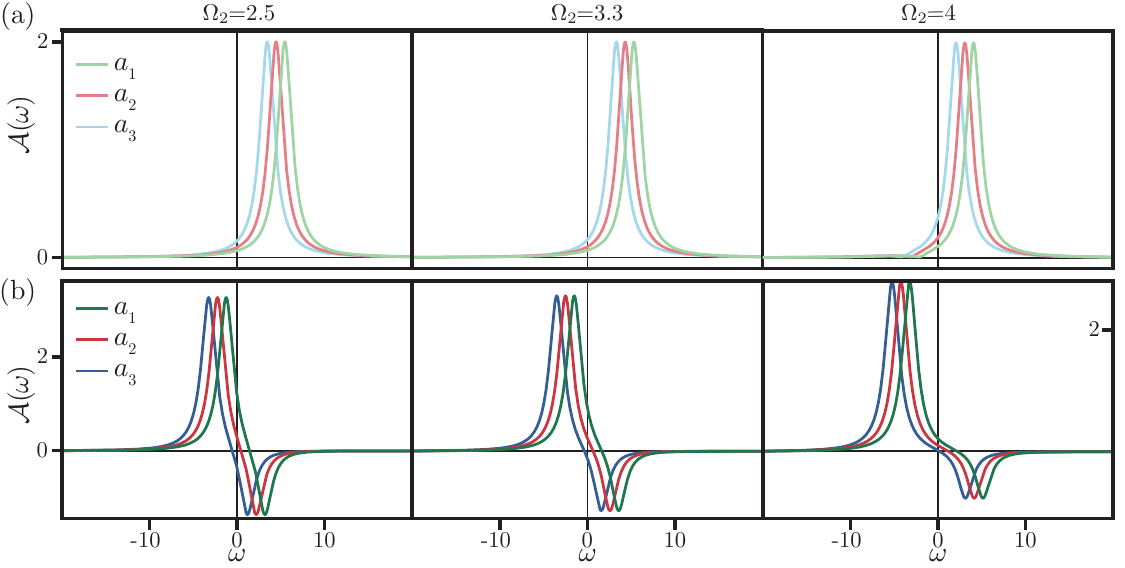}
    \centering
    \caption{Spectral function, $\mathcal{A}(\omega)$, of the three-mode harmonic cavity uniform phases, for fixed detuning $\Delta_2=+5$ and $U_0=-1$. Green, red and blue correspond to the $1^\textrm{st}$-, $2^\textrm{nd}$- and $3^\textrm{rd}$-mode, respectively. LP phase (top) and HP (bottom) at points $S_{1,2,3}$ in Fig.~\ref{fig:phase diagram}. The LP shows no particular features whereas the HP shows the typical peak inversion characterizing a dissipation stabilized excited state.}
    \label{fig:3mode_spectral_function}
\end{figure}

Using a Keldysh action approach we readily get access to the Green's function of the system and associated dynamical observable. In Fig.~\ref{fig:3mode_spectral_function}, we compare the spectral functions, $\mathcal{A}(\omega) = -2\text{Im}[G^R(\omega)]$, of the three-modes in the LP (top) and HP (bottom) phases for different pump strengths at fixed detuning $\Delta_2=+5$, points $S_{1,2,3}$ in Fig.~\ref{fig:phase diagram}. From the top panel we see that the LP phase has the response of a ``ground state", i.e., a positive (negative) peak at positive (negative) frequencies, see Sec.~\ref{sec:closed_system_results}. We note that peaks at negative frequencies are unresolved due to the scale resolution. On the other hand, the HP phase features a peak swap with a positive (negative) peak at negative (positive) frequencies, features of a stabilized excited state [cf. Sec.~\ref{sec:closed_system_results}]. We note that the two uniform phases have different occupation of the $2^\textrm{nd}$-mode, see Fig.~\ref{fig:3-mode MF} LP and HP, nevertheless their spectral functions highlight a more profound difference between them than just a different mode occupation. Interestingly, the peak swap is also present in the response of the empty side modes and we can think of it as being in presence of a normal phase to excited normal phase transition.

\section{Beyond MF and $2^\textrm{nd}$-order Cumulants}
\label{appendix:2nd_cumulant}
Following the approach of employing cumulant expansion for exploring the many-body dynamics and phase transitions as detailed in \cite{Kubo1962,Reiter2020} we perform an extension to the second order, \textit{i.e.} writing the three-body correlations appearing in EoM of Eq.~\ref{eq:EoM 3-mode cavity} as multiplications of the two-body and single-body moments. In what follows we detail the approach for the single-mode Kerr cavity and compare the results with ones obtained from the MF and the full density matrix solution. The extension to the 3-mode case directly follows the approach here to derive the EoM for 15 independent first- and second-order correlations. 

\subsection{single-mode Kerr cavity}
For the single-mode driven cavity we have
\begin{widetext}~\label{eq:EoM <a1>}
\begin{align}
    \frac{d}{dt} \braket{\hat{a}} = & -i \left(\Delta - i\gamma \right) \braket{\hat{a}} - iU_0 \braket{\hat{a}^\dagger \hat{a}^2} - i \Omega_2 = -i \left(\Delta - i\gamma + 2 U_0 \braket{\hat{a}^\dagger \hat{a}} - 2 U_0 \braket{\hat{a}^\dagger} \braket{\hat{a}}  \right) \braket{\hat{a}} - i U_0 \braket{\hat{a}^2} \braket{\hat{a}^\dagger} - i\Omega_2 \, ,
\end{align}
\end{widetext}
where we assumed a vanishing $3^\textrm{rd}$-order cumulant hence replacing the three-body correlation of $\braket{\hat{A} \hat{B} \hat{C}}$ in terms of the two-body correlation and the single-body expectation values as follow
\begin{equation*}
  \braket{\hat{A} \hat{B} \hat{C}} = \braket{\hat{A}\hat{B}}   \braket{\hat{C}} + \braket{\hat{A} \hat{C}} \braket{\hat{B}} + \braket{\hat{B} \hat{C}} \braket{\hat{A}} - 2 \braket{\hat{A}} \braket{\hat{B}} \braket{\hat{C}}
\end{equation*}
Note that the above relation is basically the same as Wick's theorem results for the Gaussian states.

Therefore, we have
\begin{equation}
    \braket{\hat{a}^\dagger \hat{a}^2} = 2\left( \braket{\hat{a}^\dagger \hat{a}} - \braket{\hat{a}^\dagger} \braket{\hat{a}} \right) \braket{\hat{a}} + \braket{\hat{a}^2} \braket{\hat{a}^\dagger} \, .
\end{equation}
As can be seen from Eq.~\ref{eq:EoM <a1>}, the two-body correlations are needed to describe the dynamics of $\braket{\hat{a}}$.
\begin{widetext}~\label{eq:EoM <a1>}
\begin{align}
    \frac{d}{dt} \braket{\hat{a}^\dagger \hat{a}} & =  - 2\gamma \braket{\hat{a}^\dagger \hat{a}} + i \Omega_2 \left(\braket{\hat{a}} - \braket{\hat{a}^\dagger} \right) \, , \\ \notag
    \frac{d}{dt} \braket{\hat{a}^2} & = -i2 \Omega_2 \braket{\hat{a}} - i2 \left(\Delta - i\gamma - 2.5  U_0 \right) \braket{\hat{a}^2} - i2 U_0 \braket{\hat{a}^3 \hat{a}^\dagger}\, .
\end{align}
\end{widetext}

The last term of the second equation involves the four-body correlation of $\braket{\hat{a}^3 \hat{a}^\dagger}$, rendering the sets of EoM to an non-closed set.

\begin{widetext}
\begin{align}
    \braket{\hat{A} \hat{B} \hat{C} \hat{D}} & = \braket{\hat{A} \hat{B}} \braket{\hat{C} \hat{D}}
    + \braket{\hat{A} \hat{C}} \braket{\hat{B} \hat{D}}
    + \braket{\hat{A} \hat{D}} \braket{\hat{B} \hat{C}} - 2 \braket{\hat{A}} \braket{\hat{B}} \braket{\hat{C}} \braket{\hat{D}}\, .
\end{align}
\end{widetext}

Finally, the desired 4-body correlation of the single-mode Kerr cavity reads as
\begin{equation}
    \braket{\hat{a}^3 \hat{a}^\dagger}  = 3 \braket{\hat{a} \hat{a}^\dagger} \braket{\hat{a}^2} - 2 \braket{\hat{a}}^3 \braket{\hat{a}^\dagger} \, .
\end{equation}

As can be seen, the last equation is in terms of the single-body expectation values, \textit{i.e.} $\braket{\hat{a}} , \braket{\hat{a}^\dagger}$, and two-body correlations as $\braket{\hat{a}^2} , \braket{\hat{a}^\dagger \hat{a}} $, only.

Finally, the EoM for the Gaussian approximation of the solution reads as
\begin{widetext}~\label{eq:EoM <a1>}
\begin{align}
    \frac{d}{dt} \braket{\hat{a}} & =  -i \left(\Delta - i\gamma + 2 U_0 \braket{\hat{a}^\dagger \hat{a}} - 2 U_0 \braket{\hat{a}^\dagger} \braket{\hat{a}}  \right) \braket{\hat{a}} - i U_0 \braket{\hat{a}^2} \braket{\hat{a}^\dagger} - i\Omega_2 \, , \\ \notag
     \frac{d}{dt} \braket{\hat{a}^\dagger \hat{a}} & =  - 2\gamma \braket{\hat{a}^\dagger \hat{a}} + i \Omega_2 \left(\braket{\hat{a}} - \braket{\hat{a}^\dagger} \right) \, , \\ \notag
     \frac{d}{dt} \braket{\hat{a}^2} & = -i2 \Omega_2 \braket{\hat{a}} - i2 \left(\Delta - i\gamma + 0.5 U_0 + 3 U_0 \braket{\hat{a}^\dagger \hat{a} } \right) \braket{\hat{a}^2} + i4 U_0 \braket{\hat{a}}^3 \braket{\hat{a}^\dagger}  \, .
\end{align}
\end{widetext}

\begin{figure}[htbp]
\centering
\includegraphics{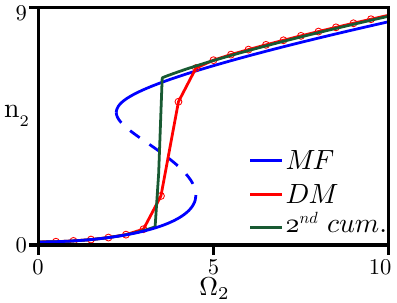}
\caption{\label{fig:cumulant approach} Comparison between MF (blue line), $2^\textrm{nd}$-cumulant (green line) \textit{i.e.} the Gaussian approximation, and the full density matrix (red line) solutions in a single-mode cavity at $\Delta = +5$. For the MF solutions, the solid and dashed lines correspond to the stable and unstable solutions, respectively. $U_0$ = -1 in all calculations.}
\end{figure}

Figure~\ref{fig:cumulant approach} compares the results of the MF-theory (blue lines) with the the full density matrix solution (red line) and the one obtained from the Gaussian approximation (green line). As can be seen the Gaussian approximation not only matches with the full density matrix solutions far away from the phase transition point but also captures the transition behavior from the LP to the HP. Specifically, the MF bistability region and the associated hysteresis behavior disappear and the two branches are connected via a rapid but continuous change. The remained discrepancy between the $2^\textrm{nd}$-cumulant results and the density matrix within the transition region is related to the non-Gaussian nature of the cavity state due to the quartic interaction. We
note that the multistable behavior does not appear in the analytic solution and the quantum solution is unique, while the semiclassical approach gives multiple dynamically
stable solutions. 

\end{document}